\documentclass[reprint,
superscriptaddress,
 amsmath,amssymb,
 aps,
prb,
]{revtex4-2}

\usepackage{graphicx}% Include figure files
\usepackage{dcolumn}% Align table columns on decimal point
\usepackage{bm}% bold math
\usepackage[utf8]{inputenc}
\usepackage[T1]{fontenc}
\usepackage{pxfonts}
\usepackage{mathptmx}
\usepackage{xfrac,nicefrac}
\usepackage{etoolbox}
\usepackage{subfigure}
\usepackage[table]{xcolor}
\graphicspath{ {Figures} } 
\usepackage[normalem]{ulem}
\usepackage{multirow}

\begin{document}

%---------------title------------------------
\title{Nuclear and magnetic structure of an epitaxial La$_{0.67}$Sr$_{0.33}$MnO$_{3}$ film using diffraction methods \\}

%------------aurthors------------------------
\author{H. Himanshu}
\affiliation{Institute Laue-Langevin, 71 avenue des Martyrs, 38000 Grenoble, France}
\affiliation{Laboratoire National des Champs Magnétiques Intenses (LNCMI-CNRS, Université Grenoble Alpes), 25 avenue des Martyrs, 38000 Grenoble, France \\}

\author{E. Rebolini}
\email[Correspondence at ]{rebolini@ill.fr}
\affiliation{Institute Laue-Langevin, 71 avenue des Martyrs, 38000 Grenoble, France \\}

\author{K. Beauvois}
\affiliation{Commissariat à l’Énergie Atomique (CEA), IRIG, MEM, MDN, Univ. Grenoble Alpes, 38000 Grenoble, France \\}

\author{S. Grenier}
\affiliation{Institut Néel (CNRS), 25 avenue des Martyrs, 38000 Grenoble, France \\}

\author{B. Mercey}
\affiliation{CRISMAT, ENSICAEN-CNRS UMR6508, 6 bd. Maréchal Juin, 14050 Caen, France \\}

\author{B. Domenges}
\altaffiliation[Deceased]{}
\affiliation{CRISMAT, ENSICAEN-CNRS UMR6508, 6 bd. Maréchal Juin, 14050 Caen, France \\}

\author{B. Ouladdiaf}
\affiliation{Institute Laue-Langevin, 71 avenue des Martyrs, 38000 Grenoble, France \\}

\author{M. B. Lepetit}
%\email[]{marie-bernadette.lepetit@neel.cnrs.fr}
\affiliation{Institute Laue-Langevin, 71 avenue des Martyrs, 38000 Grenoble, France}
\affiliation{Institut Néel (CNRS), 25 avenue des Martyrs, 38000 Grenoble, France \\}

\author{C. Simon}
\affiliation{Laboratoire National des Champs Magnétiques Intenses (LNCMI-CNRS,  Université Grenoble Alpes), 25 avenue des Martyrs, 38000 Grenoble, France \\}

\date{\today}

%-----------------Abstract------------------------------------
\begin{abstract}
We use a combination of transmission electron microscopy, X-Ray and neutron
diffraction, SQUID magnetometry and symmetry analysis to determine the
nuclear and magnetic structure of an epitaxial LSMO film, deposited on a Si
substrate. The film undergoes a magnetic ordering transition at 260~K. At
300~K, in the paramagnetic phase, the manganite film has a $I4/m$ space
group.  In the magnetic phase, SQUID and neutron diffraction results lead us
to assign a ferromagnetic spin-order to the film, associated to magnetic
moments with a slightly out-of-plane component, namely
$(3.5, 0, 0.3)\,\mu_B$. Symmetry analysis further shows that only the
$P\bar 1'$ group is compatible with such a magnetic ordering.
\end{abstract}

\maketitle

%---------------Introduction------------------------------------
\section{\label{sec:intro} Introduction \\}
Doped perovskite manganites of general formula RE$_{1-x}$AE$_{x}$MnO$_{3}$
(RE: Rare Earth, AE: Alkaline Earth) have attracted considerable interest
owing to their rich phase diagrams~\cite{1,2,3,4,5,6}. The complex interplay
of charge, lattice, spin and orbital degrees of freedom is particularly
interesting in La$_{1-x}$Sr$_{x}$MnO$_{3}$ bulk compounds. At a doping level
of $x = 0.33$ (La$_{0.67}$Sr$_{0.33}$MnO$_{3}$ - LSMO), these compounds are
found to be half-metallic ferromagnets below 370~K, and insulating in the
paramagnetic phase~\cite{7,8}. This behavior can be explained using the
double-exchange mechanism and the Jahn-Teller effect in the Mn$^{3+}$
ions~\cite{DbleExch_Zener}. In addition, colossal magneto-resistive effects
were established in these materials~\cite{10,11}. These properties make LSMO a
promising candidate for many spin-based and electronic devices. The
suitability of thin films for such applications causes a surge of interest in
miniaturization of these bulk systems.

%%%% importance of geometry in LSMO %%%%
In epitaxial thin films, the substrate imposes a strain on the system, adding
an additional degree of freedom. Due to the double-exchange mechanism, this
substrate-strain greatly modifies the film properties compared to the
bulk. As a consequence, the electronic and magnetic properties of LSMO films
can be largely manipulated by the substrate choice, the film thickness and
growth conditions~\cite{12,13,14,15,16,27}. Although the dramatic changes in
LSMO film properties have been well-studied, there is no exact
crystallographic knowledge, despite the high focus of research on correlation
between LSMO films physical properties and structural modifications with
respect to bulk~\cite{17,18,19,23,24,28}. The huge potential of these films
for technological applications thus demands a complete nuclear and magnetic
structure determination.

%%%% this paper %%%%
In this letter, we present our results obtained on a LSMO film of thickness
40.9~nm grown on a Si substrate. Such films on Si substrate are a usual
request in semiconductor industry. A buffer layer of SrTiO$_{3}$ (STO) was
deposited in-between the substrate and the film, in order to promote a better
growth of LSMO. We studied the film magnetization along different directions,
as a function of temperature and applied field, using an MPMS XL Quantum
Design SQUID magnetometer at Institut Néel. Furthermore, single crystal X-ray
diffraction (XRD at SmartLab, CEA-CNRS, Grenoble) and neutron diffraction (ND
at D10 and D10+ at ILL, Grenoble) experiments were performed on the
epitaxial film. We provide the detailed analysis for the determination of the
atomic structure and magnetic configuration of the film. 

%------------------------Experiment--------------------------------------------
\section{\label{sec:expt}Experiment\\}

\subsection{\label{sec:synthesis} Thin film synthesis}
The LSMO film was deposited using pulsed-laser deposition (PLD) on a Si substrate with in-situ Reflection High Energy Electron diffraction at CRISMAT laboratory, Caen. First, 20 unit cells of a STO buffer layer were deposited using PLD (at 650$^{\circ}$~C, in $1.6~\times 10^{-5}$~mbar pressure) on top of the already Molecular Beam Epitaxy-deposited STO. Then, the deposition of 400~\AA{} of La$_{0.67}$Sr$_{0.33}$MnO$_{3}$ (104 unit cells) was carried out, at 650$^{\circ}$~C in a pressure of $4.2 \times 10^{-4}$~mbar, and an atmosphere of $0.1\%$-volume ozone in oxygen. The deposited film was cooled in an atmosphere of  $5.9\times 10^{-3}$~mbar pressure with a mixture of ozone (7$\%$ volume) and oxygen. The cooling rate was $50^{\circ}$~C/min, after a 30 minutes annealing period in the same atmosphere. The sample surface size is $5\times 5~{\rm mm}^2$. 

\subsection{\label{sec:instruments} Instrument details}
\begin{enumerate}
\item The initial characterization of the film was performed with 
  Transmission Electron Microscopy (TEM) using an ARM200F (double corrected
  JEOL TEM, 200 kV) microscope with z dependent contrast images at the CRISMAT
  laboratory in Caen (France).
  
\item The X-ray diffraction (XRD) measurements were carried out using the
  SmartLab Rigaku diffractometer (CEA-CNRS, Grenoble, France). The
  diffractometer is equipped with a standard Cu anode source, with optional
  double bounce Ge (220) monochromator. The detector is a NaI scintillation
  point detector. The reflections  were measured using Cu
  $K_{\alpha 1}$  radiation ($\lambda = 1.5406$~\AA). The resolution
  and beam size at sample were selected using appropriate Soller slits, on both
  incident and receiving sides. The primary beam size was 0.4 mm (vertical)
  $\times$ 12 mm (horizontal).
  
\item The magnetic characterization of the film was performed using an XL Quantum
  Design SQUID magnetometer at Institut N\'eel, Grenoble (France).
  
\item Finally, single crystal neutron diffraction (ND) experiments were
  performed on the the D10/D10+ instrument at the Institute Laue-Langevin
  (Grenoble, France). D10 was used in the four-circle mode, equipped with a
  Eulerian cradle and a He-flow cryostat. The optional energy analysis mode
  was used in order to reduce the inelastic signal coming from the
  substrate. An incident wavelength of 2.36~\AA{} was selected using a
  pyrolytic graphite monochromator, and an $^3$He point detector was used to
  collect the data. The magnetic structure was further analyzed using the
  upgraded D10+ instrument.
\end{enumerate}

%------------------------Nuclear structure-------------------------------
\section{\label{sec:nuclear} Nuclear structure determination \\}

\begin{figure}[h!]
(a) \hspace*{7.5cm} \\[-0.5cm] \hspace*{0.5cm} \includegraphics[width=7cm]{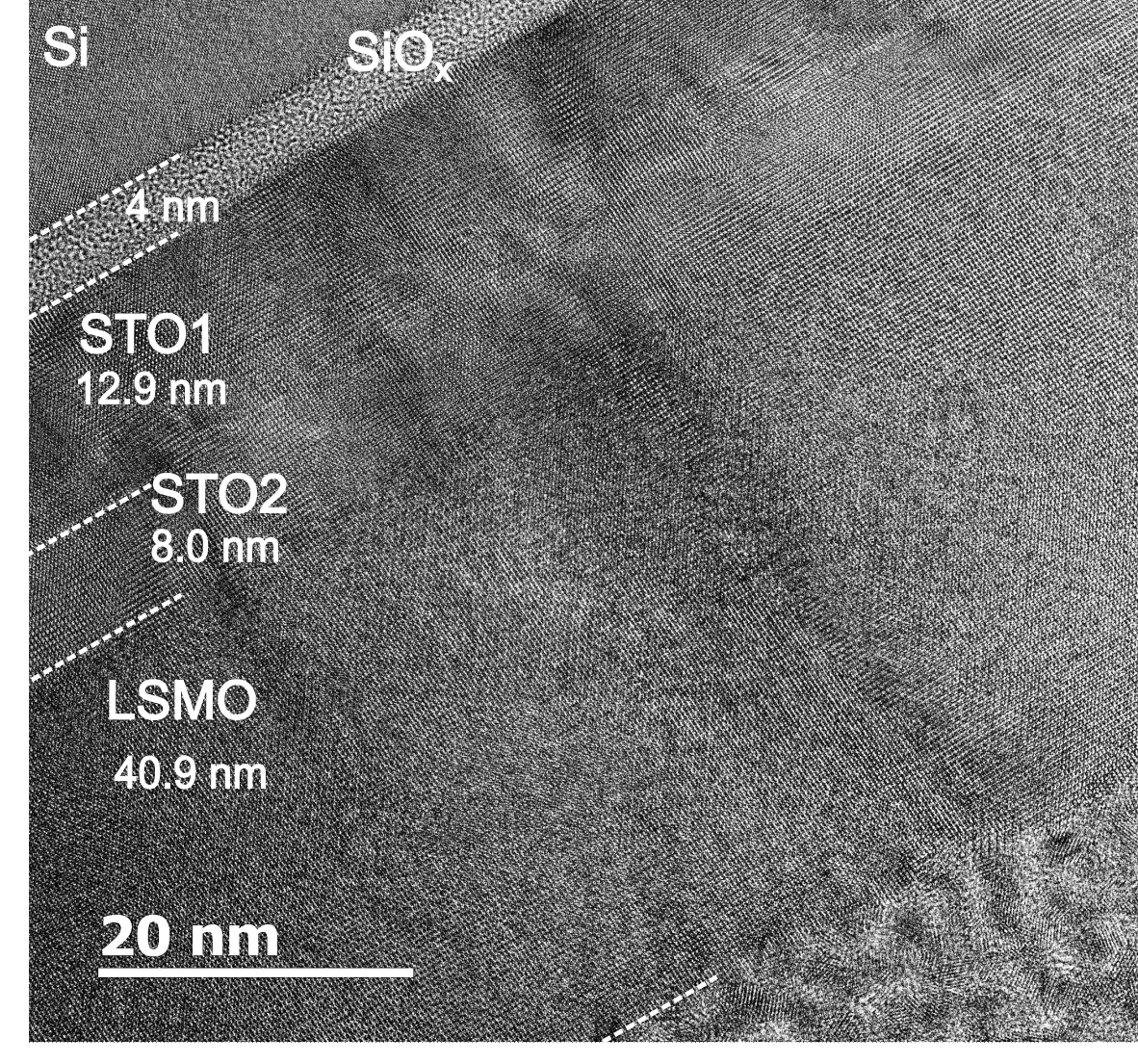} \\
(b) \hspace*{7.5cm} \\[-0.5cm] \hspace*{0.5cm} \includegraphics[width=7cm]{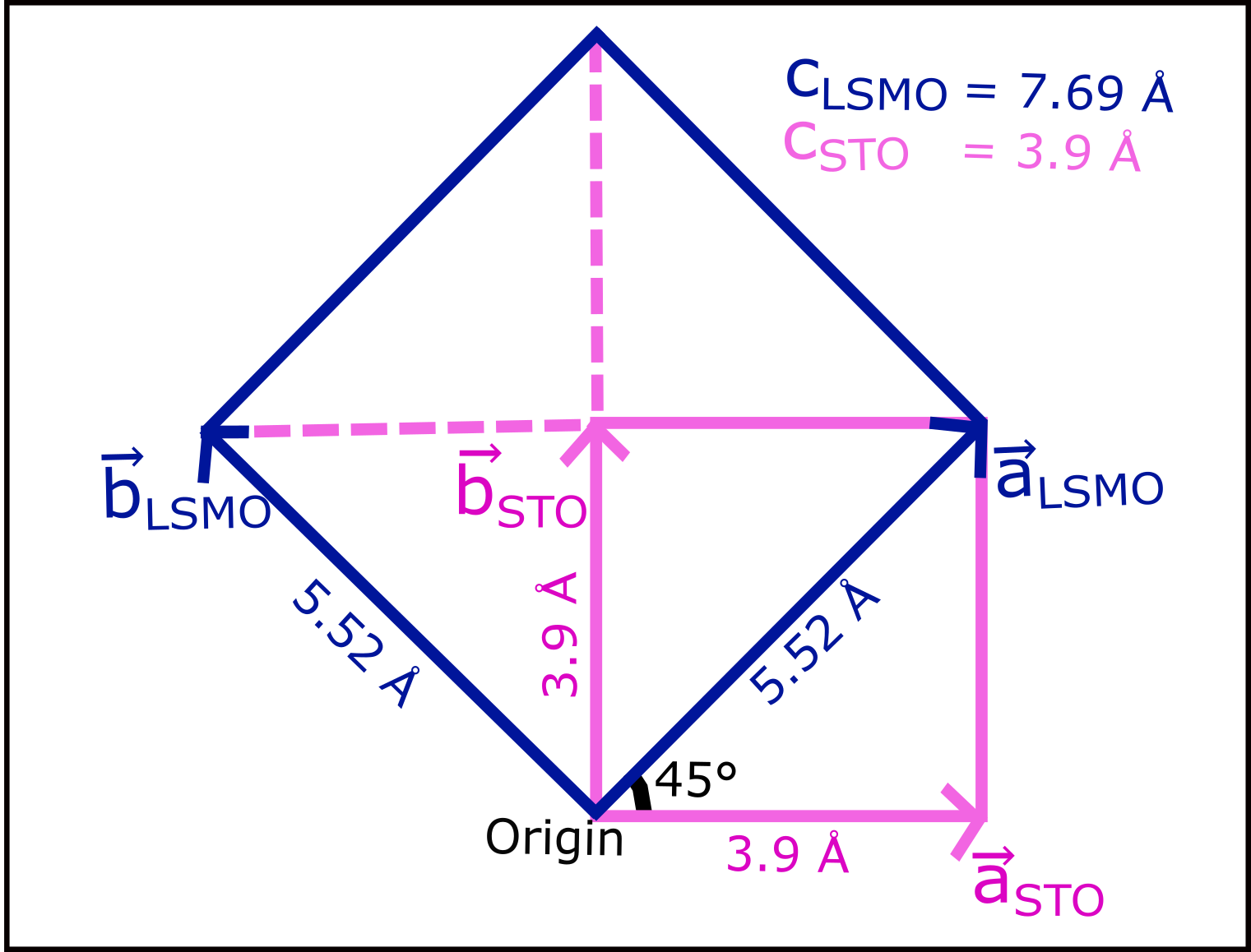}
\caption{(a) TEM image of the LSMO thin film seen along [010] direction in STO unit cell. (b) Relationship between STO and LSMO unit cells.}
\label{tem}
\end{figure}

%%%% TEM => doubling of unit cell %%%%
Figure \ref{tem}(a) shows a TEM image of
the sample, seen along $[010]$ direction, in the STO unit cell (which is cubic
with $a=b=c=3.9$~\AA). There are two domains of STO, as expected from the
deposition mechanism. The stacking of layers along the c-axis (out-of-plane direction) shows the quality of the film epitaxy. The thickness of
the LSMO film is determined to be 40.9~nm. A very thin layer (4~nm) of
silicon oxide formed in-between the substrate (Si) and STO. 

In electron diffraction experiment (cf. Appendix \ref{app-ed}), the
observation of Bragg peaks at $\nicefrac{1}{\sqrt{2}}~{\rm a}_{\text{STO}}^*$ (in-plane), and at $\nicefrac{1}{2}\,{\rm c}_{\text{STO}}^*$ (out-of-plane) in the paramagnetic phase, requires a doubling of the LSMO-film unit cell compared to the STO and the ideal-perovskite bulk-LSMO ones (see Fig.~\ref{tem}(b) for schematic representation). This results in LSMO in-plane lattice axes being rotated
by 45$^{\circ}$ compared to the STO ones, associated with a doubling in the out-of-plane direction as well ($\vec{\rm a}_{\text{LSMO}} = \vec{\rm a}_{\text{STO}} + \vec{\rm
  b}_{\text{STO}}$,
$\vec{\rm b}_{\text{LSMO}} = -\vec{\rm a}_{\text{STO}} + \vec{\rm
  b}_{\text{STO}}$, $\vec{\rm c}_{\text{LSMO}} = 2\vec{\rm
  c}_{\text{STO}}$). Henceforth, we will describe the LSMO film in this
doubled unit cell.

%%%% XRay %%%%
\begin{figure}
(a) \hspace*{7.5cm} \\[-0.5cm] \hspace*{0.5cm}  \includegraphics[width=7cm]{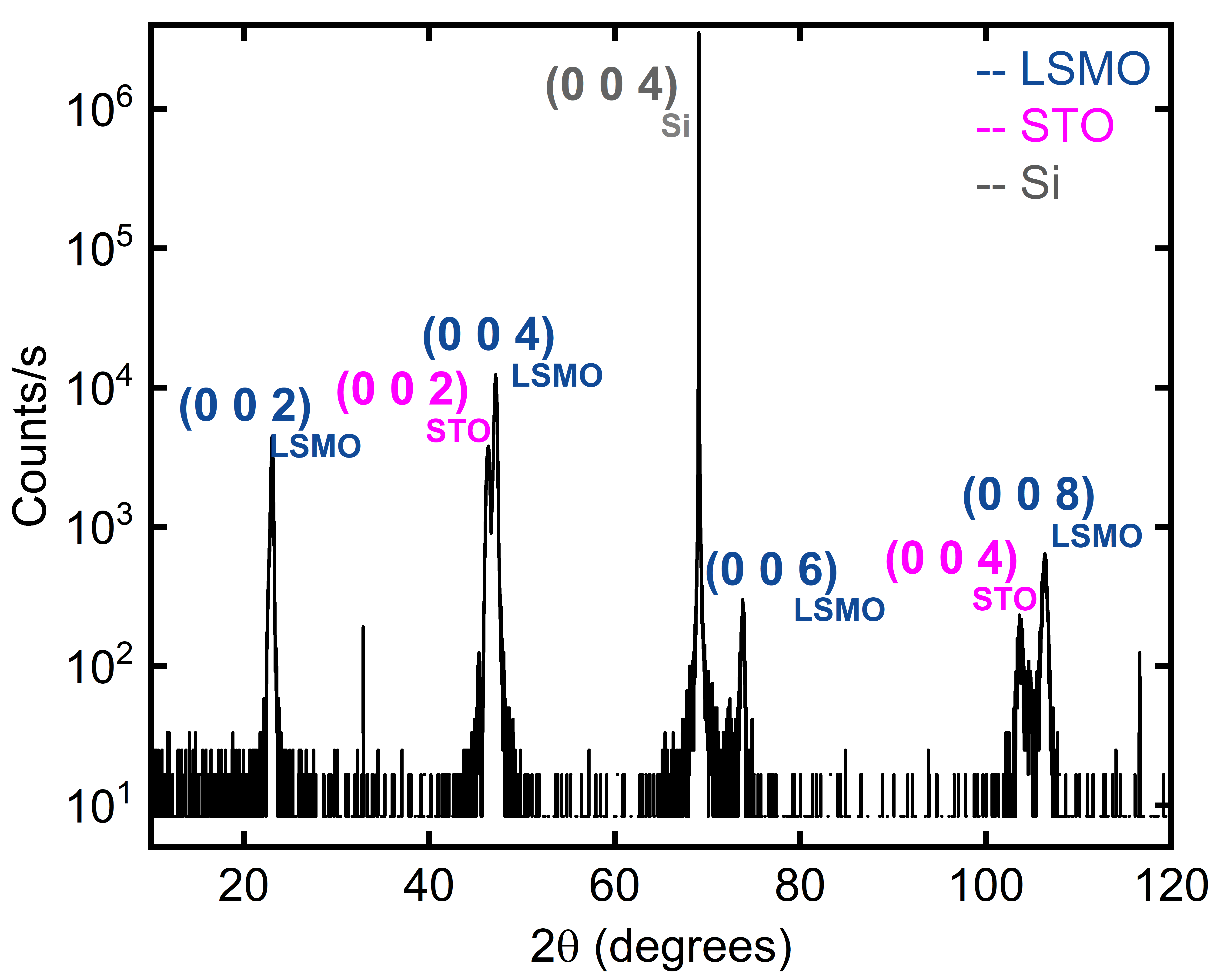}  \\
(b) \hspace*{7.5cm} \\[-0.5cm] \hspace*{0.5cm}  \includegraphics[width=7cm]{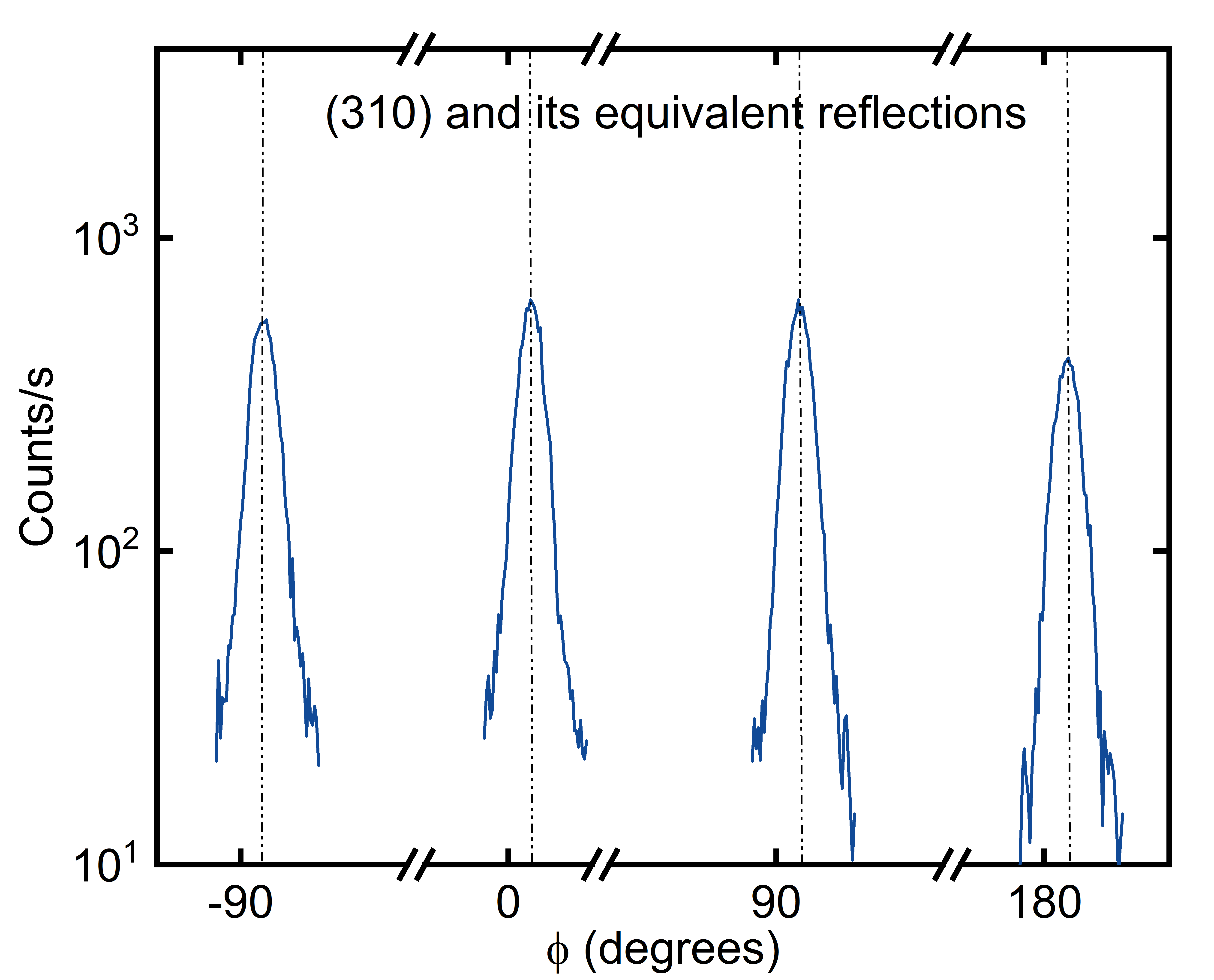} \\
(c) \hspace*{7.5cm} \\[-0.5cm] \hspace*{0.5cm}  \includegraphics[width=7cm]{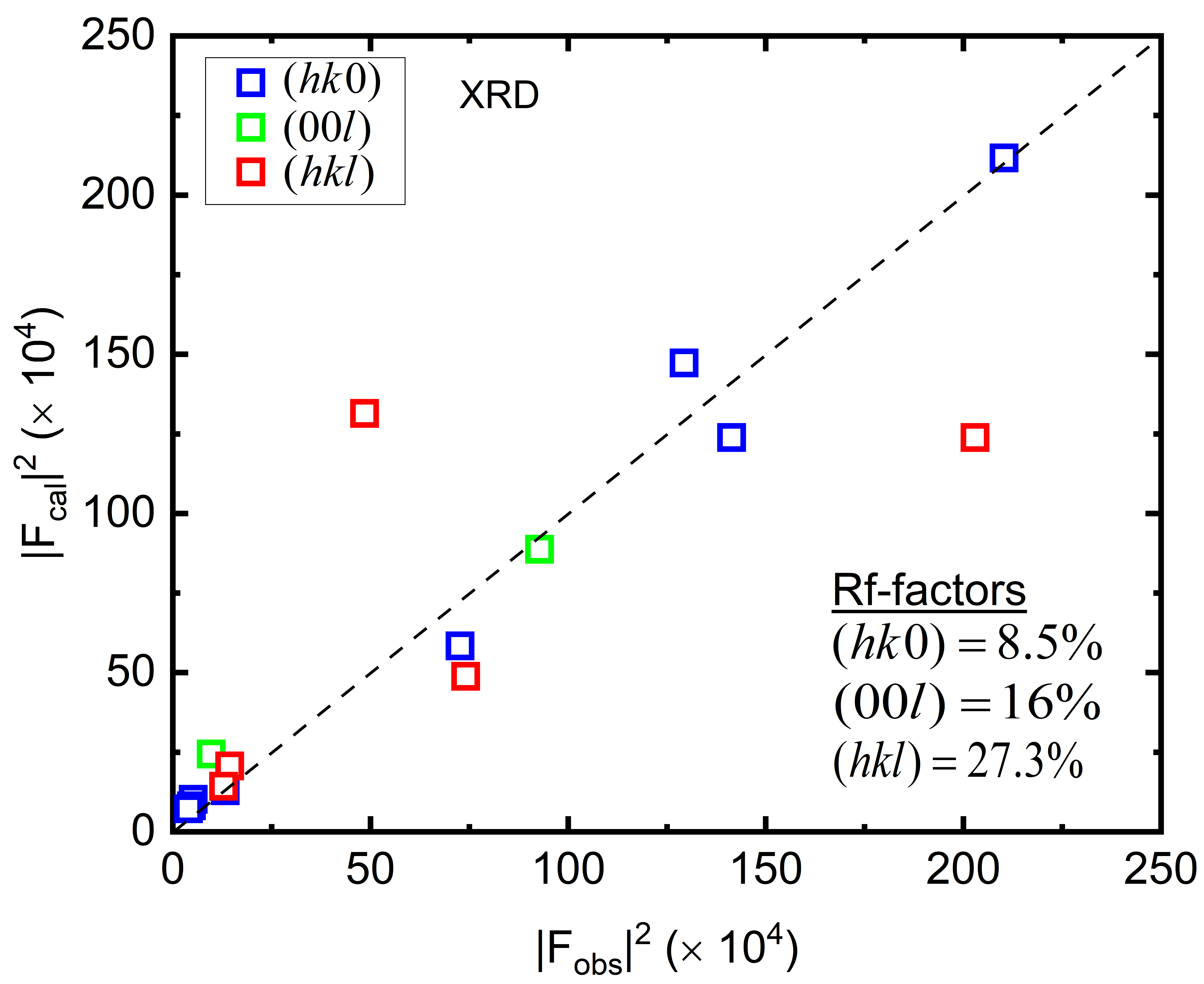}
% \hfill
\caption{XRD measurements: (a) $\omega-2\theta$ scan along (00l) depicting   peaks from Si (substrate), STO (buffer layer) and LSMO (thin film). The intensity is presented on $\log_{10}$ scale. (b) 1D
  cut of the 2D $\phi-2\theta_\chi$ (detector-sample rotation) scans of the $(310)_{\text{LSMO}}$ peak and its equivalent reflections. There is complete overlap of LSMO and STO peaks for such in-plane reflections. The intensity is presented on $\log_{10}$ scale. (c) Comparison of the observed structure factors with the calculated ones for the $I4/m$ space group. }
\label{xrd}
\end{figure}

Further insights into the crystal structure of the film were obtained using
XRD measurements at room temperature, showing that the nuclear structure of
the film follows the $I4/m$ space-group symmetry. The lattice parameters of
the body-centered cell, obtained from various XRD peak positions
---~$(0 0 \ell) \text{ and } (h k 0) \text{ families}$~--- are:
${\rm a}_{\text{LSMO}} = {\rm b}_{\text{LSMO}} = 5.522(6)$\,\AA{} and
${\rm c}_{\text{LSMO}} = 7.690(4)$\,\AA{}. The a and b lattice constants of
LSMO and STO are found to be exactly related
${\rm a}_{\text{LSMO}} = \sqrt{2}\, {\rm a}_{\text{STO}}$, which confirms that
the film is constrained by the STO layer despite the presence of a thin buffer
of silicium oxide. An $\omega-2\theta$ scan performed along $(00\ell)$ film
direction, is shown in Figure~\ref{xrd}(a). The strongest peak is
(004)$_{\text{Si}}$ from the Si substrate. The (002)$_{\text{LSMO}}$,
(004)$_{\text{LSMO}}$, (006)$_{\text{LSMO}}$, and (008)$_{\text{LSMO}}$ Bragg
reflections of the LSMO film can be seen clearly. The peaks from the STO layer
are very close to the film peaks, but can be distinguished for the (002) and
(004) reflections, thanks to the small thickness of the STO buffer layer. This
confirms that the growth of the film is along c-axis for both STO and LSMO.

The in-plane reflections ($\ell=0$) of the STO and LSMO were found in complete overlap. This reaffirms the fact that the film and the STO layer follow the same constraints and in-plane symmetry. In addition, Figure~\ref{xrd}(b) shows a cut of the 2D scans along $\phi$, corresponding to the maximum of the (310)$_{\text{LSMO}}$ peak. These well-centered equivalent peaks at consecutive $90^{\circ}$ rotation of the sample indicate the presence of an in-plane 4-fold symmetry in the film. 

In total, 56 reflections (15 independent) were measured using two different instrumental set-ups: in-plane and out-of-plane. Several corrections were applied to the measured counts/s, namely the normalization with respect to film area under diffraction, the Lorentz factor and the polarization~\cite{22}. The integrated intensities were calculated by performing numerical addition of counts/s using a homemade python script (cf. supplementary material). 

The refinement of the nuclear structure was performed with the FullProf package~\cite{20}, using least-square fitting to compare observed and calculated structure factors. The reliability of the fit was measured using Rf-factor~\cite{21}.
The data were divided into 3 sets $(hk0)$, $(00\ell)$, and the general one $(hk\ell)$, which were refined separately as we used different experimental set-ups. In the $I4/m$ (No.~79) space-group, the obtained Rf-factors are: Rf-$(hk0) = 8.5~\%$, Rf-$(00\ell) = 16~\%$, Rf-general$ = 27.3~\%$. Figure~\ref{xrd}(c) shows the comparison between the squares of calculated and observed structure factors ($|F_{calc}|^2$ and $|F_{obs}|^2$) for different reflections. The relatively high value of Rf-general can be attributed to a combination of the La/Sr disorder present in the system and the strain-relaxation of the film. This is argued because the out-of-plane peaks are asymmetric, with a long tail on the small $2\theta$ side, indicating increase in the $c$ lattice parameter (cf. supplementary material).

\begin{table}
  \caption{\label{pos-xrd} Atomic positions and Wyckoff sites of the LSMO film
    in the $I4/m$ space group (a=b=5.52~\AA, c=7.69~\AA). }
\begin{ruledtabular}
\begin{tabular}{lcl}
Atom & Wyckoff site & Fractional coordinates\\
\hline
 Mn1 & 2a & $(0, 0, 0)$ \\
 Mn2 & 2b & $(0, 0, \nicefrac{1}{2})$\\
 La/Sr & 4d & $(0, \nicefrac{1}{2}, \nicefrac{1}{4})$ \\
 O1 & 4e & $(0, 0, \nicefrac{1}{4})$ \\
 O2 & 8h & $(\nicefrac{1}{4}, \nicefrac{1}{4}, 0)$\\
\end{tabular}
\end{ruledtabular}
\end{table}

Due to limited number of measured reflections, it is important to note that only the scale factors were refined during the fitting process. Removing the mirror plane perpendicular to the c-axis ($I4$ space group) gives the possibility to optimize the z-positions of all the atoms, but the refinement is unstable due to the too-large number of parameters. To overcome this problem, we ran a 3D loop varying the z-positions of atoms and rotating the in-plane oxygen octahedra through a homemade Python script with the FullProf software. It was found that the minima of the 3 Rf-factors is at negligible distortions from the original structure. There is therefore no indication that the mirror plane is absent and hence, the best possible fit of the nuclear structure data is with $I4/m$ space group. The atomic positions are provided in Table~\ref{pos-xrd}. One should however remember that  the resolution of such XRD measurements on thin films is not good-enough to identify small distortions possibly leading to a symmetry breaking. 

%------------------------Magnetic structure-------------------------------
\section{\label{sec:magnetic} Magnetic structure determination \\}
%%%% SQUID %%%%
\subsection{\label{sec:squid} SQUID}
\begin{figure}[h!]
(a) \hspace*{7.5cm} \\[-0.5cm] \hspace*{0.5cm} \includegraphics[width=7cm]{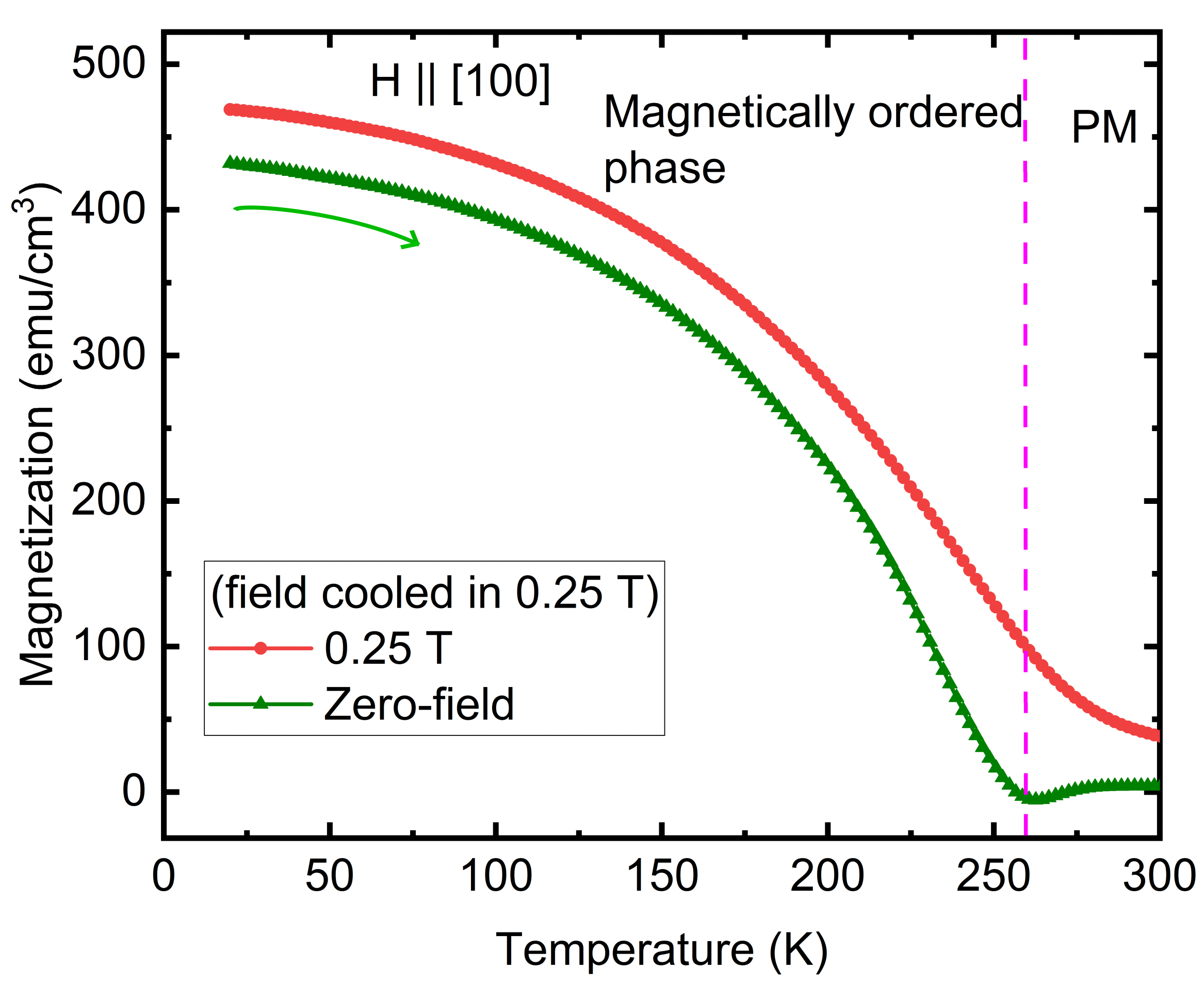} \\
(b) \hspace*{7.5cm} \\[-0.5cm] \hspace*{0.5cm} \includegraphics[width=7cm]{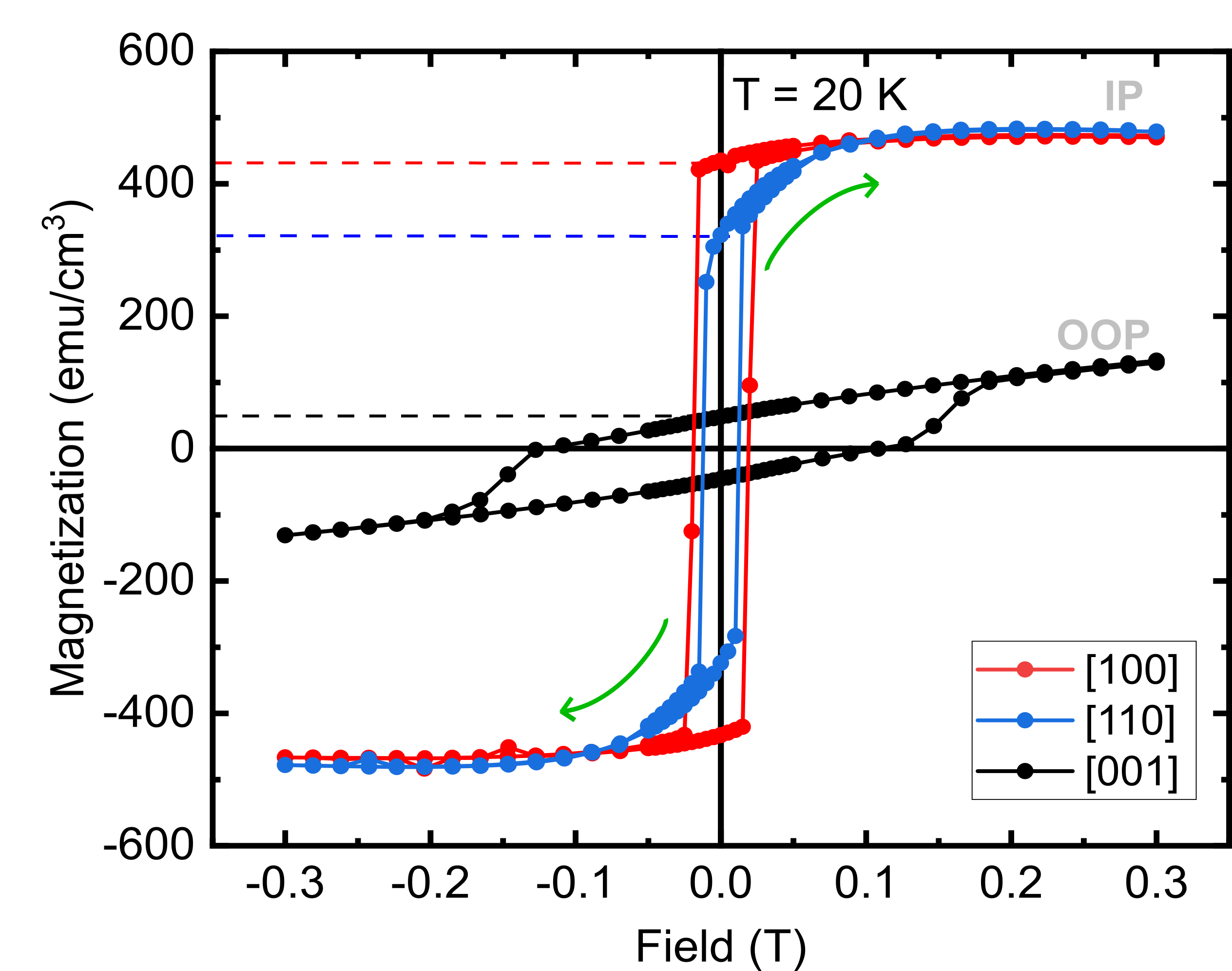}
\caption{SQUID magnetic measurements: (a) Temperature dependence of
  magnetization for applied field along in-plane $[100]$, in 0.25~T and
  zero-field (field cooled). The dashed line near 260~K marks the transition
  from paramagnetic (PM) to magnetically ordered phase. (b) Magnetic hysteresis
  loops at 20~K for field applied up to 0.3~T along $[100]$, $[110]$ and $[001]$
  film directions.}
\label{squid}
\end{figure}

Macroscopic magnetic properties of the film were investigated by performing
SQUID measurements in various applied magnetic fields, over a wide range of
temperature. We first measured the susceptibility in the paramagnetic phase and
found it very small. Since the Si susceptibility is known to be temperature
independent, the magnetic contribution of the substrate can be considered as
negligible in a 0.3~T field. Figure \ref{squid}(a) shows the magnetization
measured as a function of increasing temperature, after cooling in an in-plane
0.25~T field. The two curves (in the absence of field and under an applied
field of 0.25~T) clearly show the magnetic ordering transition at 260~K.

Figure~\ref{squid}(b) shows the magnetic hysteresis at 20~K along the 3 film
directions: $[100]$, $[110]$, and $[001]$.  The moments along the $[100]$ direction saturate in a
$\approx~0.1$~T field. The coercive field ($H_c$) is $0.02$~T, indicating a soft
magnetic nature. The shape of the hysteresis shows that the moments switch
from the $[100]$ to the $[\bar 100]$ direction instantaneously, when a field equal to
$H_c$ is applied in the reverse direction. Such a behavior is typical from a
mono-domain ferromagnetic (FM) spin arrangement.  The $[110]$ hysteresis loop
exhibits the same saturation and instantaneous switching as the $[100]$ one,
while $H_c$ is about half ($0.01$~T).  The spontaneous magnetization in the
$[100]$ direction ($\mu_a$) is very close to its saturation value, indicating
that $[100]$ is the easy-axis direction.  The fact that the spontaneous
magnetization in the $[110]$ direction is  $\sfrac{1}{\sqrt{2}}\; \mu_a$
can be interpreted as a rotation of the moments from the applied field
direction toward the easy-axis direction, when the field along $[110]$ is
reduced below 0.1~T.  The slope in the region from 0 to 0.1~T (blue curve in
Figure \ref{squid}(b)) is the result of this moment rotation.

The out-of-plane hysteresis behavior is slightly different. The moments are
not saturated at 0.3~T and $H_c$ is higher ($\approx$ 0.13 T), suggesting a
strong magnetic anisotropy. Ultimately, the moments are completely aligned
along $[001]$ in a field of $\approx$ 2 T (see supplementary material). When the applied field is reduced to zero, the non-zero magnetization
along $[001]$ shows that the moments are not completely in-plane but have an
out-of-plane component ($\mu_c$) (see Fig.~\ref{squid}(b)). The ratio between
$\mu_c$ and $\mu_a$ is about 0.1, meaning that the out-of-plane tilt of the
moments is small. Unlike in-plane, the out-of-plane hysteresis loops shows
that the moments switch from the $[001]$ to the $[00\bar 1]$ direction in a
continuous, linear way.

From these observations, one can deduce that the moments components in zero
field are $(\mu_a,0,\mu_c)$, the $[100]$ in-plane components being
ferromagnetically aligned.  Based on the shape of out-of-plane hysteresis
curves, two different scenarios can be proposed. Either the moments are
ordered anti-ferromagnetically along the $\vec {\rm c}$ axis, with a small
uncompensated magnetization of the $2a$ and $2b$ sites given by the
$\mu_c$/$\mu_a$ ratio, or the system is completely FM in nature with moment
($\mu_a$, 0, $\mu_c$).  Let us note here that a magnetic moment with both an
in-plane and out-of plane component is incompatible with magnetic atoms
located at a crystallographic site with a rotation axis along $\vec {\rm c}$,
as the 2a and 2b Mn Wyckoff-sites of the $I4/m$ group, obtained from room
temperature XRD (see Table~\ref{pos-xrd}).

%%%% ND %%%%
\subsection{\label{sec:nd} Neutron diffraction}
In order to distinguish between these two models and determine the moment
size, single crystal ND measurements were performed.
The film was aligned using the substrate Si peak positions and several
reflections were measured using $\omega-2\theta$ and $\omega$ scans. All the
measured peaks were integrated using Gaussian function fitting. The lattice
parameters obtained from ND at 300~K are as follows:
${\rm a}_{\text{LSMO}} = {\rm b}_{\text{LSMO}} = 5.51\pm 0.04$~\AA{} and
${\rm c}_{\text{LSMO}} = 7.71\pm 0.02$~\AA. These values are in good agreement
with the XRD results.

\begin{figure}
(a) \hspace*{7.5cm} \\[-0.5cm] \hspace*{0.5cm} \includegraphics[width=7cm]{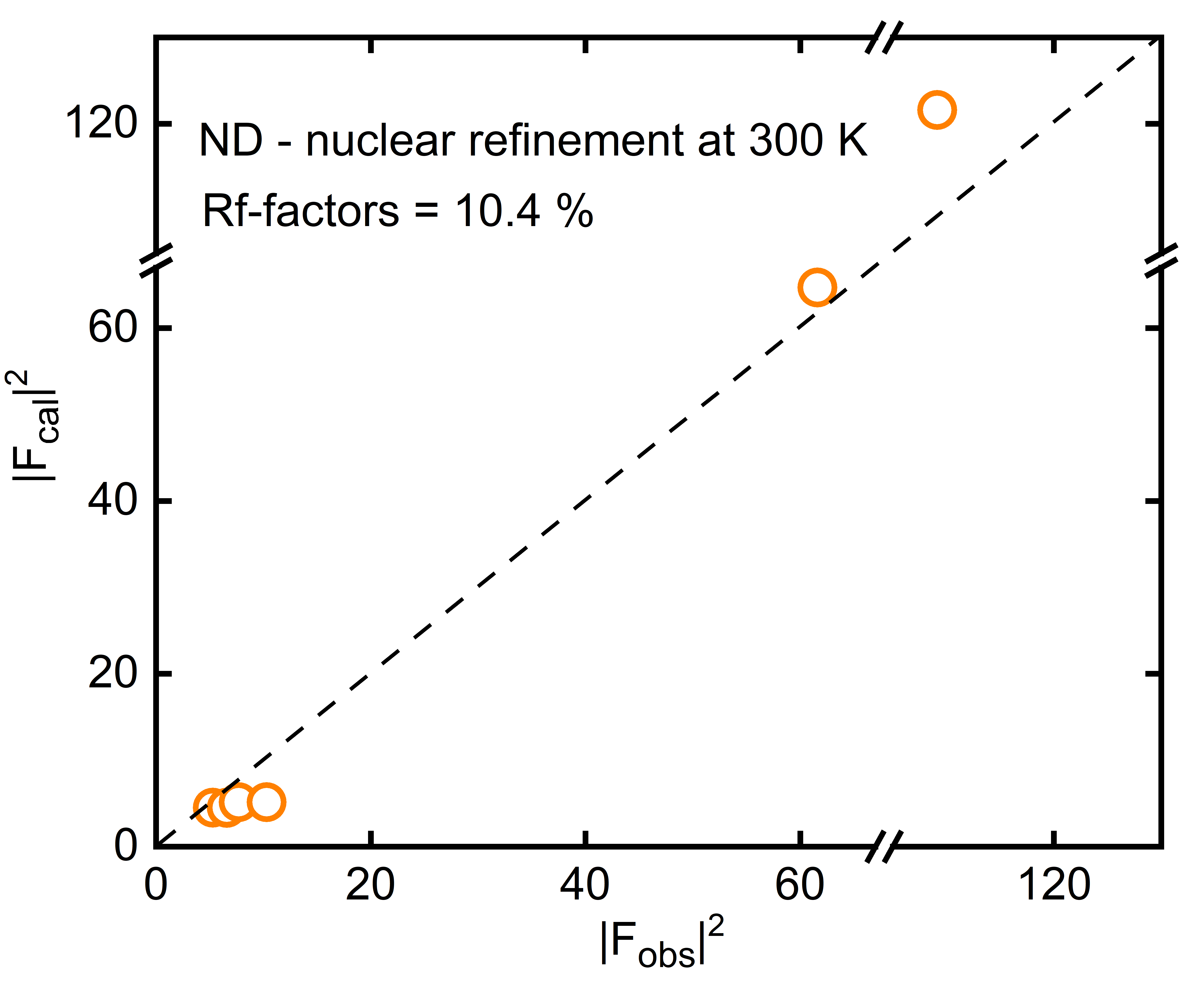} \\
(b) \hspace*{7.5cm} \\[-0.5cm] \hspace*{0.5cm} \includegraphics[width=7cm]{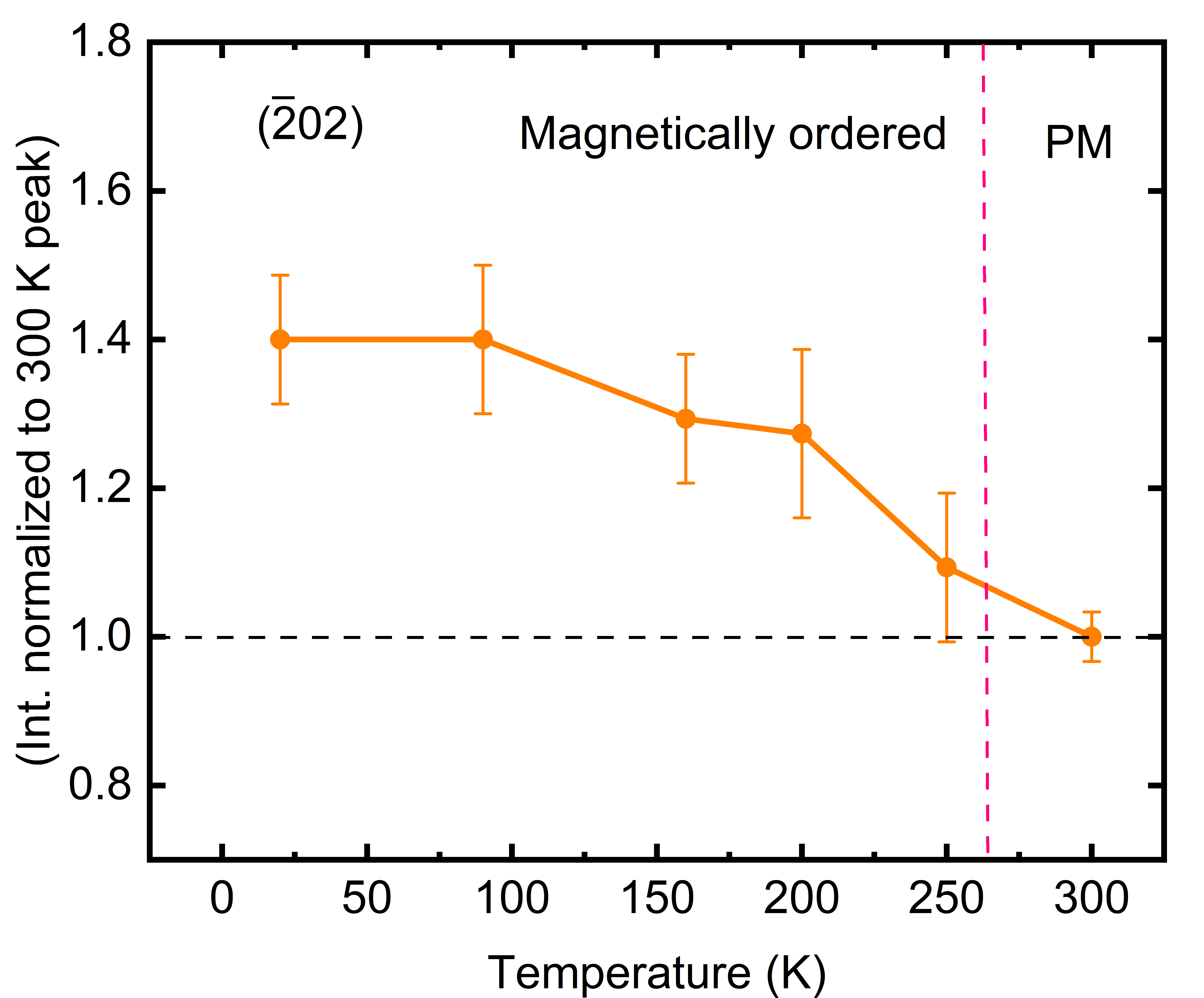} \\
(c) \hspace*{7.5cm} \\[-0.5cm] \hspace*{0.5cm} \includegraphics[width=7cm]{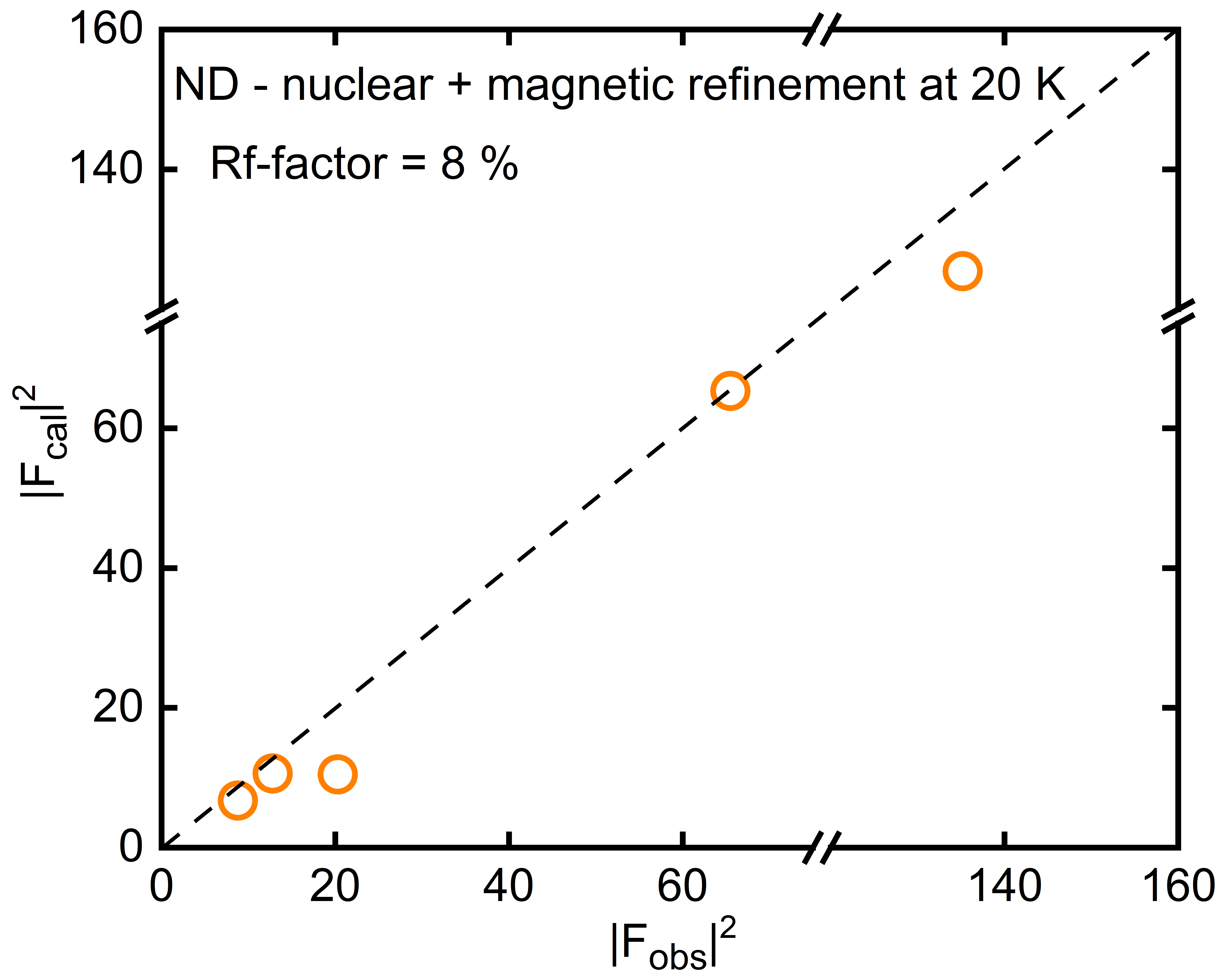}
\caption{Single crystal neutron diffraction measurements using D10. (a) Fit of
  the nuclear data at 300~K with $I4/m$ space group. (b) Temperature
  dependence of the integrated intensity of the $(\bar 202)$ peak, in the
  range 20~-~300~K, measured using $\omega-2\theta$ scans. (c) Fit of the ND
  data at 20~K with nuclear and magnetic refinement.}
\label{nd}
\end{figure}

The peaks measured at 300~K (paramagnetic phase of the LSMO film)
were fitted with the $I4/m$ nuclear space group using FullProf. The
comparison of the observed and calculated structure factors is shown in Figure
\ref{nd}(a). For the nuclear data, only the scale factor was refined and it is
in agreement with the $I4/m$ space group (Rf-factor = 10.4\%). The
temperature dependence of a few peaks intensities was followed in the range
20~-~300~K using  $\omega-2\theta$ scans. The magnetic intensity
appears on existing reflexions and is responsible for the peaks increase below
250~K. As a consequence the long-range magnetic ordering propagation vector is
$\vec q=(000)$. The magnetic contribution is shown for the $(\bar 202)$ reflection
in Figure~\ref{nd}(b) and is consistent with the SQUID results discussed
above.

From SQUID measurements, we know that the in-plane moments are aligned ferromagnetically but need to establish the out-of-plane ordering using ND. An initial set of 5 reflections (containing both nuclear and magnetic contributions) was measured on D10 in the magnetically ordered phase, at 20~K. Unfortunately, this set was not enough to answer the out-of-plane coupling question. Measurement of some extra reflections became possible thanks to the upgrade of D10 into D10+. On D10+, we measured peaks with pure magnetic content, that are absent for an FM structure. The set  includes the (201), (111) and (101) peaks. Our theoretical magnetic intensity calculations (using FullProf) showed that these peaks should be  sufficient to distinguish between different antiferromagnetic (AFM) models on the out-of-plane component. 

\begin{figure} [h!]
(a) \hspace*{7.5cm} \\[-0.5cm] \hspace*{0.5cm} \includegraphics[width=7cm]{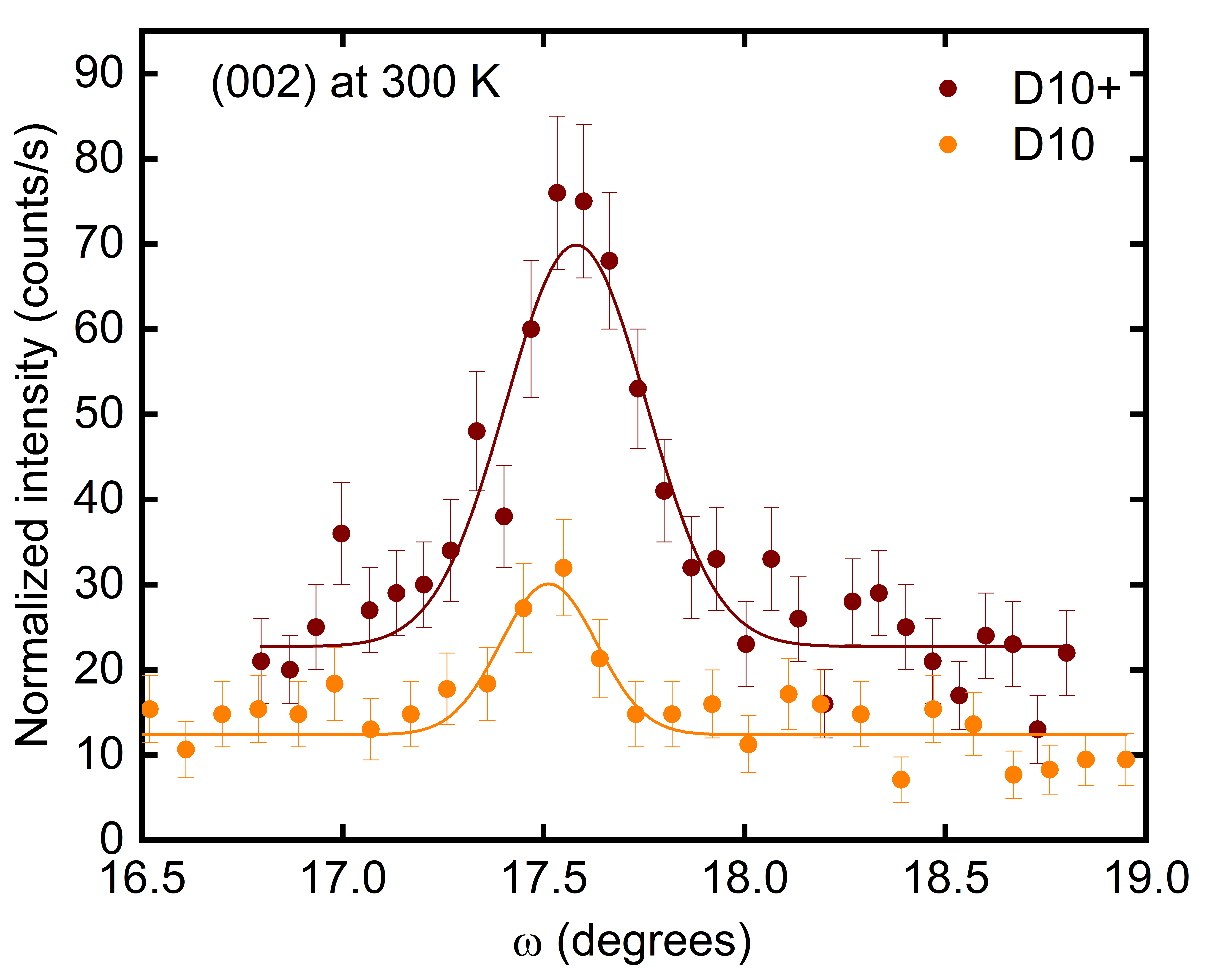} \\
(b) \hspace*{7.5cm} \\[-0.5cm] \hspace*{0.5cm} \includegraphics[width=7cm]{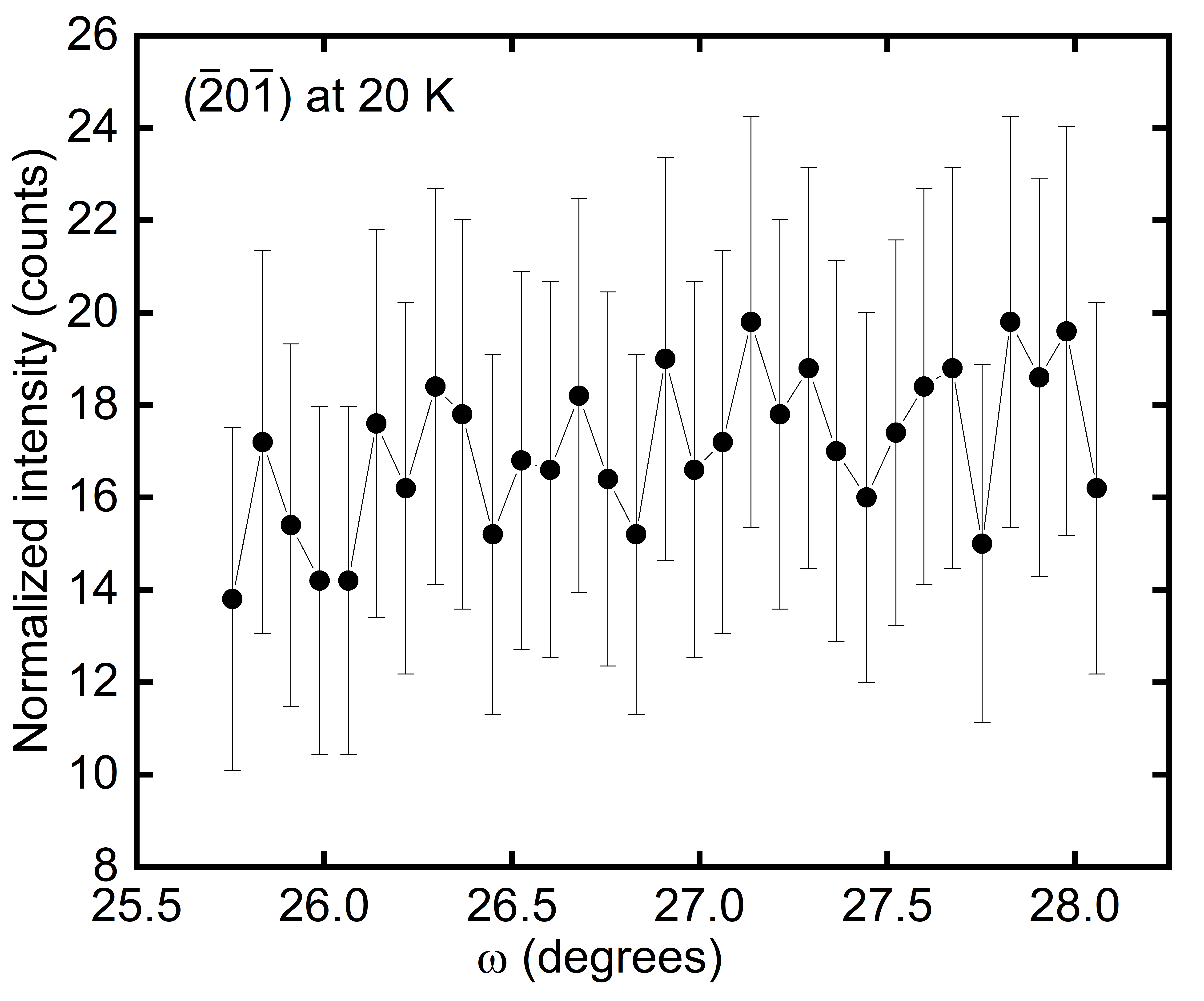}
\caption{Single-crystal neutron diffraction measurement on ILL-D10+. (a) Comparison of the (002) peak intensities at 300~K, measured using D10 and upgraded D10+. (b) Purely magnetic AFM $(\bar 20\bar 1)$ measured at 20~K.}
\label{d10+}
\end{figure}

Figure~\ref{d10+}(a) displays the (002) peak measured at 300~K with both D10
and D10+, with normalized measurement time. As can be clearly seen, the D10+
upgrade provides a large increase in intensity ($\sim 2.5$). This is a
remarkable achievement, that will allow significant improvements to studies
like this one where we need information on weak peaks. This is particularly
important as we did not detect any intensity on the set of peaks selected to
decipher the AFM ordering of the moment out-of-plane component. The $(\bar 20\bar 1)$
reflection is shown as matter of example on Fig.~\ref{d10+}(b). We can thus
conclude that the ordering is completely FM in nature.

The $\mu_a$ and $\mu_c$ components of the moments were then evaluated by the
fitting of the 20~K data-set in FullProf. Since the $\vec{a}$ and $\vec{b}$
axes are supposed to be equivalent in the film, two magnetic domains
$(\mu_a,0,\mu_c)$ and $(0, \mu_a, \mu_c)$ were used with 50 $\%$ occupancy. In
the magnetic refinement, the nuclear structure was described in the $P\bar 1$
space group (see Appendix \ref{app-symm} for detailed symmetry analysis) and
the scale factor taken from the 300~K data. The $\mu_a/\mu_c$ ratio was
optimized while keeping the norm of the moment fixed at its theoretical value
of 3.7$~\mu_B$ per Mn ($0.67 \mu_{\rm Mn^{3+}}+0.33 \mu_{\rm Mn^{4+}}$). While
it was not possible to accurately determine the ratio, the best Rf-factor was
for $0.1\le\mu_a/\mu_c\le0.3$, in agreement with the SQUID experiment.  The
SQUID $\mu_c/\mu_a=0.1$ ratio gives an Rf-factor of $8\,\%$ as shown in
Figure~\ref{nd}(c). The Mn moment for this model is $(3.5,0,0.3)\mu_B$. It
corresponds to an out-of-plane tilt of $\simeq 6^{\circ}$. Let us remember
that in bulk LSMO, the magnetic order is also FM, but the easy-axis is along
$[111]$~\cite{JONKER1950337}, whereas in ultra-thin films, less then 3~nm thick,
there is a co-existence of FM and AFM phases~\cite{16,25}. In this regard, we
can say that our film, with a 40.9~nm thickness, is in an intermediate regime,
not fully constrained, but not yet fully relaxed.

%%%Symmetry analysis%%%
\subsection{\label{sec:symm} Symmetry analysis}
Symmetry analysis shows that the resolved magnetic structure is not compatible
with the $I4/m$ crystallographic group. In fact
such a FM arrangement of moments, with both in-plane and out-of-plane
components, is only compatible with the $P\bar 1$ space group ($P \bar 1'$
magnetic group) with Mn at any Wyckoff site (see Appendix \ref{app-symm} for detailed calculations). By continuity with the $I4/m$
group found in XRD we can use a $I \bar 1'$ group with the
Mn atoms in $2a$ and $2b$.  It is clear here that diffraction
techniques do not provide data of high-enough quality on such thin films, to detect
small distortions in the structure and thus weak symmetry breaking.

%------------------------Conclusion-------------------------------
\section{\label{sec:conclusion} Conclusion \\}
In summary, we synthesized a high quality epitaxial LSMO thin-film 40.9~nm
thick, on a Si substrate, with a buffer layer of STO. The thin thickness of
the buffer layer allowed us to distinguish the film diffraction peaks from
those of STO. We showed that single crystal diffraction measurements can be
performed on such highly epitaxial films using both X-rays and neutrons. The
300~K nuclear structure was refined in the $I4/m$ space group using XRD
data, and corroborated using the neutron diffraction ones.  Below 260~K the
film undergoes a paramagnetic to ferromagnetic transition, the magnetic
structure being resolved using a combination of SQUID and ND measurements. The
Mn magnetic moment were found to have and essentially in-plane orientation but
a weak out-of plane component and were refined to be $(3.5, 0,
0.3)\,\mu_B$. Considering the fact that the bulk magnetic moment are along the
$[111]$ direction and that very-thin films present a coexistence of FM and AFM
phases, one can speculate that at such thickness we are in the regime where
the in-plane symmetry constrains imposed by the substrate start to
relax~\cite{p:LSMO}, while the in-plane unit-cell lengths are still frozen.
One can also think that the amplitude of the moment out-of-plane component can
be controlled by the film thickness, and that at a somewhat thinner thickness
one may have FM films, with fully in-plane magnetic moments and $I4/m$
symmetry.

Such studies combining X-Ray and neutron diffraction, SQUID measurements and
symmetry analysis allow considerable insight into the structural and magnetic
properties of very thin films. \\

The authors have no conflicts to disclose.

The data that support the findings of this study are available from the corresponding author upon reasonable request and will be available from \url{data.ill.eu} at \url{https://doi.ill.fr/10.5291/ILL-DATA.5-54-371} after an embargo period of 3 years.

%---------------------------Appendix-----------------------------------
\appendix

%%%Electron diffraction%%%
\section{\label{app-ed} Electron diffraction results: doubling of the unit cell}
\begin{figure}[h!]
\centering
\includegraphics[width=7cm]{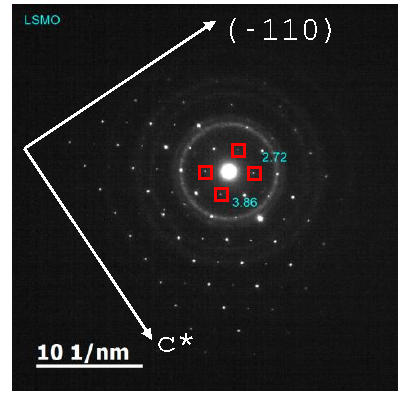}
\caption{Electron diffraction peaks measured on the LSMO film. Red squares
  highlight the peaks requiring a unit cell increase (direction (-110) and c$^*$ are in
  the  STO unit-cell notations)}
\label{ed}
\end{figure}

Figure~\ref{ed} displays the LSMO film electron diffraction peaks in the reciprocal
space labeled in the STO basis set.  The peaks with the smallest components
along each directions (squared in red on the figure) correspond to
$$ \pm(-\nicefrac{1}{2},\,\nicefrac{1}{2},\,\pm\nicefrac{1}{2})_{\text{STO}} $$
as $2.72\text{ \AA{}}=\nicefrac{1}{\sqrt{2}}\; {\rm a}_{\text{STO}}$.  The
  existence of such a diffraction peak in the paramagnetic phase directly
  requires a doubling of the LSMO film unit cell both the in-plane and
  out-of-plane, as compared to the STO one. It results for the LSMO film
  lattice parameter
  \begin{eqnarray}
    \vec{\rm a}_{\text{LSMO}} = \vec{\rm a}_{\text{STO}} + \vec{\rm b}_{\text{STO}} \\
    \vec{\rm b}_{\text{LSMO}} = -\vec{\rm a}_{\text{STO}} + \vec{\rm b}_{\text{STO}} \\
    \vec{\rm c}_{\text{LSMO}} = 2\, \vec{\rm c}_{\text{STO}}
\label{eqn-cell}
\end{eqnarray}

%%%Symmetry analysis%%%
\section{\label{app-symm} Space group symmetry analysis}
We have shown that the Mn magnetic moments have both an in-plane and an
out-of-plane component.

Group theory tells us that a magnetic atom, located on a site with a rotation
axis (or on a mirror plane), can only have a magnetic moment 
either along this axis or in the perpendicular plane (either in the plane or along
the perpendicular axis).
As a result the Mn atoms of the film cannot be located at a Wyckoff site with
a rotation around the $\vec {\rm c}$ axis or a mirror plane perpendicular to
the latter. This simple statement strongly constrain the possible space groups
of the film.

Indeed in the paramagnetic phase space group~: $I4/m$, the Mn atoms can be
located either on a Wyckoff's site of 4 degeneracy ($4c$, $4d$, $4e$, $4f$) or
on two Wyckoff's of degeneracy 2 ($2a$ and $2b$). All those sites however have
a two- or four-fold rotation axis along $\vec {\rm c}$ (see
Table~\ref{sites}), invalidating $I4/m$ as a possible LSMO space group for the
magnetic phase.

Keeping the I centering the subgroups of $I4/m$ with $k_{\text{index}}=1$ are
$I\bar 4$, $I4$, $I2/m$, $Im$, $I2$, they other one being only $P\bar1$ and
$P1$. Again, one can easily see from Table~\ref{sites} that the only
possibilities compatible with the existence of both an in-plane and
out-of-plane component of the magnetization are $I2/m$ with the Mn ions at the
$4e$ or $4f$ locations, the $I2$ with the Mn ions at the $4c$ location or the
$Im$ with the Mn ions at the $4b$ location.

\begin{table}[h!]
  \caption{\label{sites} Different possible Wyckoff's sites for the Mn atom, in the $I4/m$ 
    group and its subgroups keeping the body-centered unit cell.}
  \centering
  \begin{tabular}{lcr|lcr}
    \hline \hline \rule{0pt}{2.5ex}{\hfill}
    SG & Possible Mn  & \multicolumn{1}{c|}{Site}&
        SG & Possible Mn  & \multicolumn{1}{c}{Site}\\
       & Wyckoff's site & symmetry &
       & Wyckoff's site & symmetry \\
    \hline \rule{0pt}{2.5ex}{\hfill}
    \multirow{4}{*}{$I4/m$} 
       & \textit{2a} + \textit{2b} & ..4/m 
             & \multirow{6}{*}{$I2/m$} &  \textit{2a} + \textit{2b} & ..2/m\\
       & \textit{4c} & ..2/m & &  \textit{2c} + \textit{2d} & ..2/m\\
       & \textit{4d} & ..4 & & \textit{4e} & -1 \\
       & \textit{4e} & ..4 & & \textit{4f} & -1 \\
    \cline{1-3} \rule{0pt}{2.5ex}{\hfill}
     \multirow{4}{*}{$I\bar 4$}  &  \textit{2a} + \textit{2b} & ..-4 & & \textit{4g} & ..2 \\
       & \textit{2c} + \textit{2d} & ..-4     & & \textit{4h} & ..2 \\
       & \textit{4e} & ..2 & & \textit{4i} & ..m \\
        \cline{4-6} \rule{0pt}{2.5ex}{\hfill}
       & \textit{4f} & ..2 &  \multirow{2}{*}{$I2$} & \textit{2a} + \textit{2b} & ..2  \\
    \cline{1-3} \rule{0pt}{2.5ex}{\hfill} 
    \multirow{2}{*}{$I4$} 
       & \textit{2a} + \textit{2a} & ..4 & & \textit{4c} & 1  \\
    \cline{4-6} \rule{0pt}{2.5ex}{\hfill}
       & \textit{4b} & ..2 & \multirow{2}{*}{$Im$} & \textit{2a} + \textit{2a} & ..m\\
       &&&& \textit{4b} & 1 \\
    \hline \hline
  \end{tabular}
\end{table}

Let us thus analyze those possibilities in view of the neutrons scattering
results.

\subsection{Mn on $4e$ or $4f$ sites in the  $I2/m'$ magnetic group}
The magnetic space group associated with the $I2/m$ crystallographic group is
$I2/m'$ if one considers that only the CPT($=i'$) symmetry operation is respected in our space-time.

The symmetry operations of the $I\,1\,1\,2/m'$ groups thus transform the atoms and their magnetic moment has follows \\
\begin{table}[h]
  \centering
\begin{tabular}[t]{c@{}|rrr@{\quad}|rrr}
  \hline \hline
  Sym. Op. & \multicolumn{3}{c|}{Frac. atom. coord.} & \multicolumn{3}{c}{Magn. moment} \\
  \hline 
  $1$     & $x$& $y$& $z$ & $\mu_x$ & $\mu_y$ & $\mu_z$\\
  $2_{0,0,z}$ &$-x$&$-y$& $z$ &$-\mu_x$ &$-\mu_y$ & $\mu_z$\\
  $\bar 1'$    &$-x$&$-y$&$-z$ &$-\mu_x$ &$-\mu_y$ &$-\mu_z$\\
  $m'$  & $x$& $y$&$-z$  & $\mu_x$ &$ \mu_y$ &$-\mu_z$\\
  \hline
  $t_{(\frac{1}{2}, \frac{1}{2}, 0)}$ &$x+\frac{1}{2}$&$y+\frac{1}{2}$&$z+\frac{1}{2}$
                          &   $\mu_x$ & $\mu_y$ & $\mu_z$\\
  $t\circ 2_{0,0,z}$&$-x+\frac{1}{2}$&$-y+\frac{1}{2}$&$z+\frac{1}{2}$
                          & $-\mu_x$ &$-\mu_y$ & $\mu_z$\\
  $t\circ \bar 1'$&$-x+\frac{1}{2}$&$-y+\frac{1}{2}$&$-z+\frac{1}{2}$
                          &$-\mu_x$ &$-\mu_y$ &$-\mu_z$\\
  $t\circ m'$&$x+\frac{1}{2}$&$y+\frac{1}{2}$&$-z+\frac{1}{2}$
                          & $\mu_x$ &$ \mu_y$ &$-\mu_z$\\
  \hline \hline 
\end{tabular} 
  \caption{Action of the symmetry operations in the $I2/m'$ group.}
  \label{tab:I2sm}
\end{table}

The $4e$ Wickoff's sites correspond to
$(\nicefrac{1}{4}, \nicefrac{1}{4}, \nicefrac{1}{4})$,
$(\nicefrac{3}{4}, \nicefrac{3}{4}, \nicefrac{1}{4})$,
$(\nicefrac{3}{4}, \nicefrac{3}{4}, \nicefrac{3}{4})$ and
$(\nicefrac{1}{4}, \nicefrac{1}{4}, \nicefrac{3}{4})$. They are thus invariant
under the $t\circ \bar 1'$ transformation, as should be their magnetic
moment. This provides conditions for the irreducible representation (irrep) to
which the magnetic structure belongs~; indeed it has to have a -1 character on
the $t\circ \bar 1'$ symmetry operation. This is the case for the $A_u$, $B_u$
where the $\bar 1'$ symmetry operation is associated with a -1 character, and
the $A'_g$ and $B'_g$ irreps where it is the partial translation
$t_{(\frac{1}{2}, \frac{1}{2}, 0)}$ that is associated with a -1 character.
\begin{table}[h]
  \centering
  \begin{tabular}[t]{c|ccc|rrr}
    \hline \hline
    Irrep &   \multicolumn{3}{c|}{Frac. atom. coord.} & \multicolumn{3}{c}{Magn. moment} \\
    \hline
    \multirow{4}{*}{$A_u$}
          &$\nicefrac{1}{4}$ & $\nicefrac{1}{4}$ & $\nicefrac{1}{4}$ & $\mu_x$ & $\mu_y$ & $\mu_z$\\
          &$\nicefrac{3}{4}$ & $\nicefrac{3}{4}$ & $\nicefrac{1}{4}$ & $-\mu_x$ & $-\mu_y$ & $\mu_z$\\
          &$\nicefrac{3}{4}$ & $\nicefrac{3}{4}$ & $\nicefrac{3}{4}$ & $\mu_x$ & $\mu_y$ & $\mu_z$\\
          &$\nicefrac{1}{4}$ & $\nicefrac{1}{4}$ & $\nicefrac{3}{4}$ & $-\mu_x$ & $-\mu_y$ & $\mu_z$\\
    \hline
    \multirow{4}{*}{$B_u$}
          &$\nicefrac{1}{4}$ & $\nicefrac{1}{4}$ & $\nicefrac{1}{4}$ & $\mu_x$ & $\mu_y$ & $\mu_z$\\
          &$\nicefrac{3}{4}$ & $\nicefrac{3}{4}$ & $\nicefrac{1}{4}$ & $\mu_x$ & $\mu_y$ & $-\mu_z$\\
          &$\nicefrac{3}{4}$ & $\nicefrac{3}{4}$ & $\nicefrac{3}{4}$ & $\mu_x$ & $\mu_y$ & $\mu_z$\\
          &$\nicefrac{1}{4}$ & $\nicefrac{1}{4}$ & $\nicefrac{3}{4}$ & $\mu_x$ & $\mu_y$ & $-\mu_z$\\
    \hline
    \multirow{4}{*}{$A'_g$}
          &$\nicefrac{1}{4}$ & $\nicefrac{1}{4}$ & $\nicefrac{1}{4}$ & $\mu_x$ & $\mu_y$ & $\mu_z$\\
          &$\nicefrac{3}{4}$ & $\nicefrac{3}{4}$ & $\nicefrac{1}{4}$ & $-\mu_x$ & $-\mu_y$ & $\mu_z$\\
          &$\nicefrac{3}{4}$ & $\nicefrac{3}{4}$ & $\nicefrac{3}{4}$ & $-\mu_x$ & $-\mu_y$ & $-\mu_z$\\
          &$\nicefrac{1}{4}$ & $\nicefrac{1}{4}$ & $\nicefrac{3}{4}$ & $\mu_x$ & $\mu_y$ & $-\mu_z$\\
    \hline
    \multirow{4}{*}{$B'_g$}
          &$\nicefrac{1}{4}$ & $\nicefrac{1}{4}$ & $\nicefrac{1}{4}$ & $\mu_x$ & $\mu_y$ & $\mu_z$\\
          &$\nicefrac{3}{4}$ & $\nicefrac{3}{4}$ & $\nicefrac{1}{4}$ & $\mu_x$ & $\mu_y$ & $-\mu_z$\\
          &$\nicefrac{3}{4}$ & $\nicefrac{3}{4}$ & $\nicefrac{3}{4}$ & $-\mu_x$ & $-\mu_y$ & $-\mu_z$\\
          &$\nicefrac{1}{4}$ & $\nicefrac{1}{4}$ & $\nicefrac{3}{4}$ & $-\mu_x$ & $-\mu_y$ & $\mu_z$\\
    \hline \hline
  \end{tabular}
  \caption{Magnetic moment associated with the possible irreps for the $4e$ Mn sites in the $I2/m'$ group.}
   \label{tab:I2sm-4e}
\end{table}

One sees immediately from Table~\ref{tab:I2sm-4e} that none of the allowed
irreps are associated with the FM ordering found in neutron diffraction and
SQUID experiments. Indeed, according to Table~\ref{tab:I2sm-4e}, a FM ordering
would require either pure in-plane or pure out-of-plane magnetic moments, while
both those hypotheses are incompatible with the experimental results. 

The $4f$ exhibit a behavior similar to the $4e$ one. As a consequence the
$I2/m$ crystallographic group cannot be the film space group and one has to
consider the other alternatives.

\subsection{Mn on $4c$  sites in the  $I2$ magnetic group}
The magnetic space group associated with the $I2$ crystallographic group is
$I2$ if one considers that only the CPT($=i'$) symmetry operation is respected in our space-time.

The symmetry operations of the $I\,1\,1\,2$ group thus transform the atoms and their magnetic moment has follows \\
\begin{table}[h]
  \centering
\begin{tabular}[t]{c@{}|rrr@{\quad}|rrr}
  \hline \hline
  Sym. Op. & \multicolumn{3}{c|}{Frac. atom. coord.} & \multicolumn{3}{c}{Magn. moment} \\
  \hline 
  $1$     & $x$& $y$& $z$ & $\mu_x$ & $\mu_y$ & $\mu_z$\\
  $2_{0,0,z}$ &$-x$&$-y$& $z$ &$-\mu_x$ &$-\mu_y$ & $\mu_z$\\
  \hline
  $t_{(\frac{1}{2}, \frac{1}{2}, 0)}$ &$x+\frac{1}{2}$&$y+\frac{1}{2}$&$z+\frac{1}{2}$
                          &   $\mu_x$ & $\mu_y$ & $\mu_z$\\
  $t\circ 2_{0,0,z}$&$-x+\frac{1}{2}$&$-y+\frac{1}{2}$&$z+\frac{1}{2}$
                          & $-\mu_x$ &$-\mu_y$ & $\mu_z$\\
  \hline \hline 
\end{tabular} 
  \caption{Action of the symmetry operations in the $I2$ group.}
  \label{tab:I2}
\end{table}

The $4c$ Wickoff's sites correspond to
$(\nicefrac{1}{4}, \nicefrac{1}{4}, \nicefrac{1}{4})$,
$(\nicefrac{3}{4}, \nicefrac{3}{4}, \nicefrac{1}{4})$,
$(\nicefrac{3}{4}, \nicefrac{3}{4}, \nicefrac{3}{4})$ and
$(\nicefrac{1}{4}, \nicefrac{1}{4}, \nicefrac{3}{4})$.  There are four irreps
in this group (see Table~\ref{tab:charI2}), but it is easy to see that a FM
ordering is again incompatible in this group with magnetic moment with both
in-plane and out-of-plane components.
\begin{table}[h]
  \centering
  \begin{tabular}{c|rr|rr}
    \hline   \hline
    $I2$       & E & $2_{0,0,z}$ & $t_{(\frac{1}{2}, \frac{1}{2}, 0)}$ &  $t\circ 2_{0,0,z}$  \\
    \hline 
    A             & 1 &  1 &  1 &  1   \\
    B             & 1 & -1 &  1 & -1   \\
    \hline
    A'            & 1 &  1 & -1 & -1  \\
    B'            & 1 & -1 & -1 &  1  \\
    \hline   \hline
  \end{tabular}
  \caption{Character table of the $I2$ group.}
  \label{tab:charI2}
\end{table}

Let us see the next hypothesis.

\subsection{Mn on $4b$  sites in the  $Im$ magnetic group}
The magnetic space group associated with the $Im$ crystallographic group is
$Im'$ if one considers that only the CPT($=i'$) symmetry operation is respected in our space-time.

The symmetry operations of the $I\,1\,1\,m'$ group thus transform the atoms and their magnetic moment has follows \\
\begin{table}[h]
  \centering
  \begin{tabular}[t]{c@{}|rrr@{\quad}|rrr}
    \hline \hline
    Sym. Op. & \multicolumn{3}{c|}{Frac. atom. coord.} & \multicolumn{3}{c}{Magn. moment} \\
    \hline 
    $1$     & $x$& $y$& $z$ & ~$\mu_x$ & ~$\mu_y$ & ~$\mu_z$\\
    $m'$    & $x$& $y$&$-z$ & ~$\mu_x$ & ~$\mu_y$ & $-\mu_z$\\
    \hline 
    $t_{(\frac{1}{2}, \frac{1}{2}, 0)}$ &$x+\frac{1}{2}$&$y+\frac{1}{2}$&$z+\frac{1}{2}$
                 &   $\mu_x$ & $\mu_y$ & $\mu_z$\\
    $t\circ m'$  & $x+\frac{1}{2}$&$y+\frac{1}{2}$&$-z+\frac{1}{2}$
                 & $\mu_x$ &$\mu_y$ & $-\mu_z$\\
    \hline \hline 
  \end{tabular} 
  \caption{Action of the symmetry operations in the $Im'$ group.}
  \label{tab:Im}
\end{table}

The $4b$ Wickoff's sites correspond to
$(\nicefrac{1}{4}, \nicefrac{1}{4}, \nicefrac{1}{4})$,
$(\nicefrac{1}{4}, \nicefrac{1}{4}, \nicefrac{3}{4})$,
$(\nicefrac{3}{4}, \nicefrac{3}{4}, \nicefrac{3}{4})$ and
$(\nicefrac{3}{4}, \nicefrac{3}{4}, \nicefrac{1}{4})$.  There are four irreps
in this group (see Table~\ref{tab:charIm}), but it is easy to see that a FM
ordering is again incompatible with magnetic moment with both
in-plane and out-of-plane components.
\begin{table}[h]
  \centering
  \begin{tabular}{c|rr|rr}
    \hline   \hline
    $I2$       & E & $m'$ & $t_{(\frac{1}{2}, \frac{1}{2}, 0)}$ &  $t\circ m'$  \\
    \hline 
    A             & 1 &  1 &  1 &  1   \\
    B             & 1 & -1 &  1 & -1   \\
    \hline
    A'            & 1 &  1 & -1 & -1  \\
    B'            & 1 & -1 & -1 &  1  \\
    \hline   \hline
  \end{tabular}
  \caption{Character table of the $Im'$ group.}
  \label{tab:charIm}
\end{table}

\subsection{Conclusion on symmetry analysis}
At this stage the $I4/m$, $I2/m$, $I2$ and $Im$ crystallographic groups have
been out-ruled by the magnetic structure found in neutrons scattering and SQUID
measurements. The only possibility within the same unit cell is the $P\bar 1$
or $P_1$ groups.

The magnetic groups associated with  $P\bar 1$ is  $P\bar 1'$. It transform the magnetic moments as follow
\begin{table}[h]
  \centering
\begin{tabular}[t]{c@{}|rrr@{\quad}|rrr}
  \hline \hline
  Sym. Op. & \multicolumn{3}{c|}{Frac. atom. coord.} & \multicolumn{3}{c}{Magn. moment} \\
  \hline 
  $1$     & $x$& $y$& $z$ & $\mu_x$ & $\mu_y$ & $\mu_z$\\
  $i'$    & $-x$& $-y$&$-z$ & $-\mu_x$ & $-\mu_y$ & $-\mu_z$\\
  \hline \hline 
\end{tabular} 
  \caption{Action of the symmetry operations in the $P\bar 1'$ group.}
  \label{tab:P1}
\end{table}

The Mn  atoms must now be in two independent $2i$ Wickoff's positions, that is
$(\nicefrac{1}{4}, \nicefrac{1}{4}, \nicefrac{1}{4})$, $(\nicefrac{3}{4}, \nicefrac{3}{4}, \nicefrac{3}{4})$ for the first one
and $(\nicefrac{1}{4}, \nicefrac{1}{4}, \nicefrac{3}{4})$, $(\nicefrac{3}{4}, \nicefrac{3}{4}, \nicefrac{1}{4})$
for the second one.
One can now see that a FM order correspond to the  B irrep and that both in-plane and out-of-plane components are allowed. 
\begin{table}[h]
  \centering
  \begin{tabular}{c|rr}
    \hline   \hline
    $P\bar 1'$       & E & $i'$ \\
    \hline 
    A             & 1 &  1   \\
    B             & 1 & -1   \\
    \hline   \hline
  \end{tabular}
  \caption{Character table of the $P\bar 1'$ group.}
  \label{tab:charP1}
\end{table}

One can thus conclude that among the $I4/m'$ sub-groups only the $P\bar 1'$
group is compatible with the FM ordering found in our experimental results.

\bibliography{LSMO-film}% Produces the bibliography via BibTeX.

\end{document}

\title{Nuclear and magnetic structure of an epitaxial
  La$_{0.67}$Sr$_{0.33}$MnO$_{3}$ film using diffraction methods\\}

\author{H. Himanshu}
\affiliation{Institute Laue-Langevin, 71 avenue des
  Martyrs, 38000 Grenoble, France}
\affiliation{Laboratoire National des
  Champs Magnétiques Intenses (LNCMI-CNRS, Université Grenoble Alpes), 25
  avenue des Martyrs, 38000 Grenoble, France}

\author{E. Rebolini}
\email[]{rebolini@ill.fr}
\affiliation{Institute Laue-Langevin, 71 avenue des Martyrs, 38000 Grenoble, France}

\author{K. Beauvois}
\affiliation{Commissariat à l’Énergie Atomique (CEA), IRIG, MEM, MDN, Univ. Grenoble Alpes, 38000 Grenoble, France}

\author{S. Grenier}
\affiliation{Institut Néel (CNRS), 25 avenue des Martyrs, 38000 Grenoble, France}

\author{B. Mercey}
\affiliation{CRISMAT, ENSICAEN-CNRS UMR6508, 6 bd. Maréchal Juin, 14050 Caen, France}

\author{B. Domenges~\dag}
\affiliation{CRISMAT, ENSICAEN-CNRS UMR6508, 6 bd. Maréchal Juin, 14050 Caen, France}

\author{B. Ouladdiaf}
\affiliation{Institute Laue-Langevin, 71 avenue des Martyrs, 38000 Grenoble, France}

\author{M. B. Lepetit}
\affiliation{Institute Laue-Langevin, 71 avenue des Martyrs, 38000 Grenoble, France}
\affiliation{Institut Néel (CNRS), 25 avenue des Martyrs, 38000 Grenoble, France}

\author{C. Simon}
\affiliation{Laboratoire National des Champs Magnétiques Intenses (LNCMI-CNRS,  Université Grenoble Alpes), 25 avenue des Martyrs, 38000 Grenoble, France}

\date{\today}
% =================================================================

\begin{abstract}
 \centerline{\huge \bf Supplementary Material}
\end{abstract}

\maketitle %\maketitle must follow title, authors, abstract and \pacs

\onecolumngrid

\section{Corrections and peak integration method applied to the XRD data collected on LSMO thin film}
\subsection{Film area under diffraction}
The measured counts/s were normalized with the projection of the beam area onto the sample surface. The general strategy applied is as follows:
\begin{itemize}
\item the beam imprint onto the sample (rectangular area) is calculated for both incident and detector sides~;
\item the film area that contributes to the measured intensity is the
  intersection of the smallest area out of the two sides (incident/detector), with the sample surface~;
\item These surface areas are calculated using the equations
  \begin{eqnarray}
    \textrm{incident beam area} &=& \frac{h_{slit}}{\cos\chi} \times \frac{v_{slit}}{\sin\omega}\\
    \textrm{beam area on the detector} &=& \frac{h_{slit}}{\cos\chi} \times \frac{det_{vslit}}{\sin2\theta}
  \end{eqnarray}
  where, $h_{slit}$, $v_{slit}$ and $det_{vslit}$ are respectively the horizontal incident, vertical incident and vertical detector slit sizes (mm), $\chi$ is the angle of the out-of-plane sample rotation, $\omega$ is the angle of the  incident beam  with respect to film surface, and $2\theta$ is the angle of the detector with incident beam.
\end{itemize}

\begin{figure}
(a) \hspace*{8cm} \\[-3ex]  \hspace*{0.5cm}  \includegraphics[width=7cm]{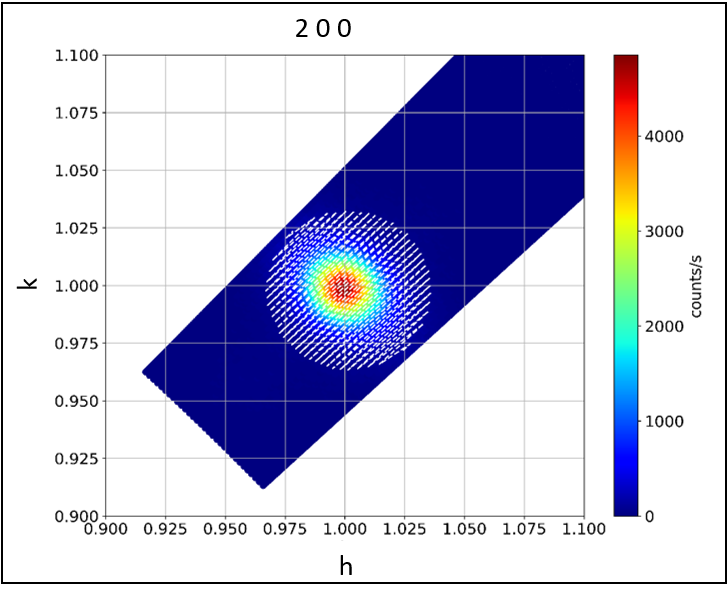}\\
(b) \hspace*{8cm} \\[-3ex]  \hspace*{0.5cm}  \includegraphics[width=7cm]{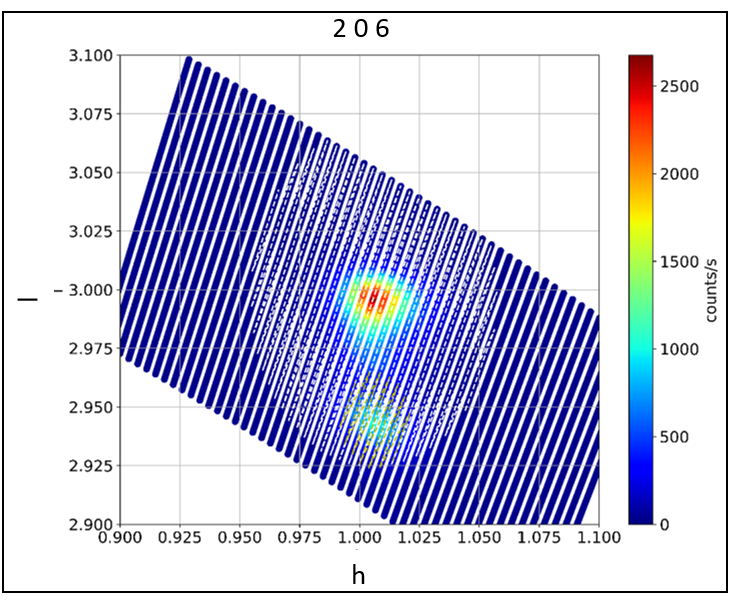}
\caption{Two of the measured XRD peaks of the LSMO film with area of
  integration shown as a dashed circle. (a) (200) peak and (b) (206) peak. The axes are labeled in the STO unit cell because the sample was aligned using the STO unit cell setting.}
  \label{xrdpeaks}
\end{figure}

\subsection{Lorentz factor and polarization}
To perform these corrections, each data point (counts/s) is divided by
$$\frac{1+A\cos^{2}2\theta}{(1+A)\sin2\theta}$$ also knows as the Lorentz
polarization factor. $A = \cos^{2}2\theta_{M}$ where $\theta_M$ is the
Bragg angle of the monochromator crystal~\cite{1,2}. In our experiment, a Ge
(220) monochromator was used for the out-of-plane geometry. Since it was a
double bounce monochromator, the final polarization correction includes $A =
\cos^{4}2\theta_{M}$ for the out-of-plane reflections.
    
\subsection{Divergence correction}
For the in-plane geometry, the reflections were measured in two separate
experiments. In order to make the two data-sets comparable, it has to be made
sure that the optical set-up is similar for both. The only difference in the
two experiments was the width of slits used for horizontal divergence on the
detector side (0.5$^{\circ}$ and 0.228$^{\circ}$). This means that more
photons were counted in the measurements with the 0.5$^{\circ}$
slit. Therefore, all the data points from this set were multiplied with the
$\sfrac{0.228}{0.5}$ factor.

\subsection{Peak integration method}
The integration strategy is to first define a radius of integration for each
peak. The final intensity is basically the sum of counts/s inside this radius
after subtracting the background. The average background is calculated from 
the data points outside the peak radius (see figure \ref{xrdpeaks}).

\subsection{Absorption correction}
The integrated intensities were corrected for the x-ray absorption by the
film. This was done using the Beer-Lambert's law
\begin{eqnarray}
  I &=& I_0 \, \exp(-\mu t)
\end{eqnarray}
Where, $I$ is the measured intensity after absorption, $\mu$ is the linear
absorption coefficient~\cite{3} and $t$ is the thickness of the material under
diffraction. For the thin film, the path traveled by X-rays inside the
material is $2t/\sin{\theta}$. Hence, the corrected intensity is given by
\begin{eqnarray}
  I_0 &=& I/\exp{(-2\mu t/ \sin{\theta})}
\end{eqnarray}

\newpage
\section{A few of the out-of-plane peaks measured using XRD on LSMO film}
\begin{figure} [h!]
(a) \hspace*{7.5cm} \\[-3ex] \hspace*{0.5cm} \includegraphics[width=0.5\linewidth]{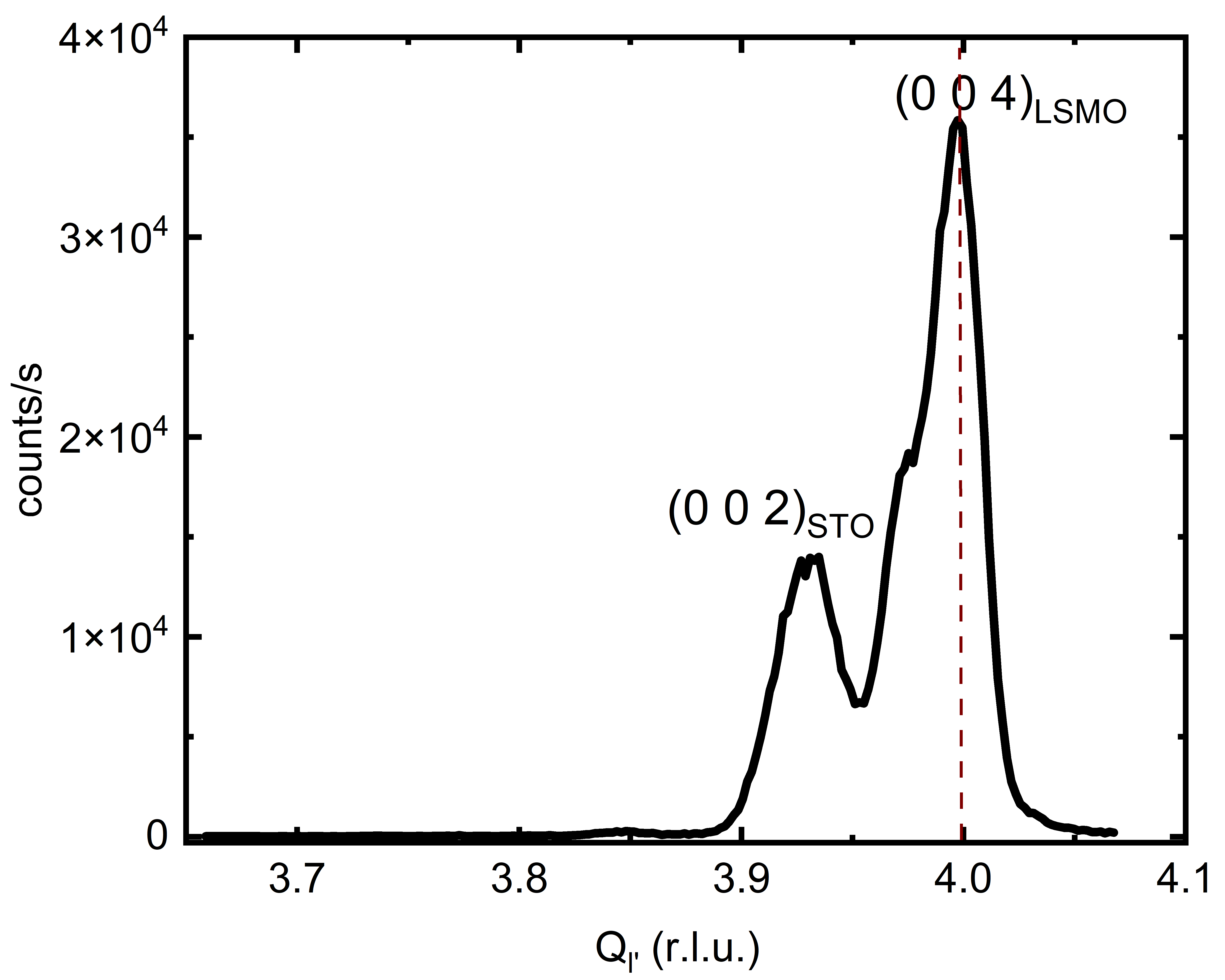}\\
(b) \hspace*{7.5cm} \\[-3ex] \hspace*{0.5cm}\includegraphics[width=0.5\linewidth]{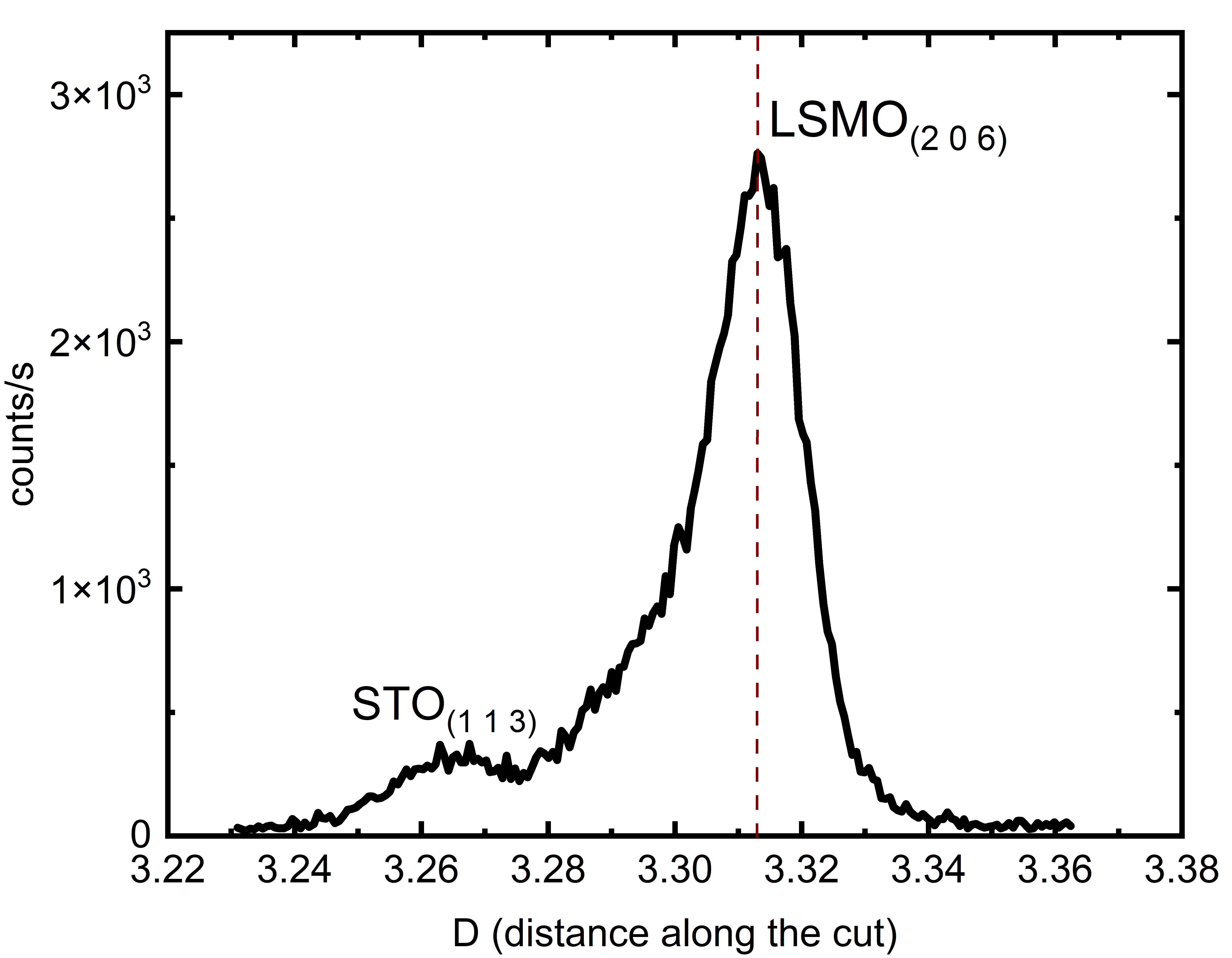}
\hfill
\caption{(a) LSMO (004) XRD peaks and (b) 1D cut of the 2D LSMO (206) scan. One should note the asymmetry seen in out-of-plane LSMO peaks.}
\label{relax}
\end{figure}

\newpage
\section{Magnetic hysteresis in 5 T field}
\begin{figure} [h!]
\centering
\includegraphics[width=0.6\linewidth]{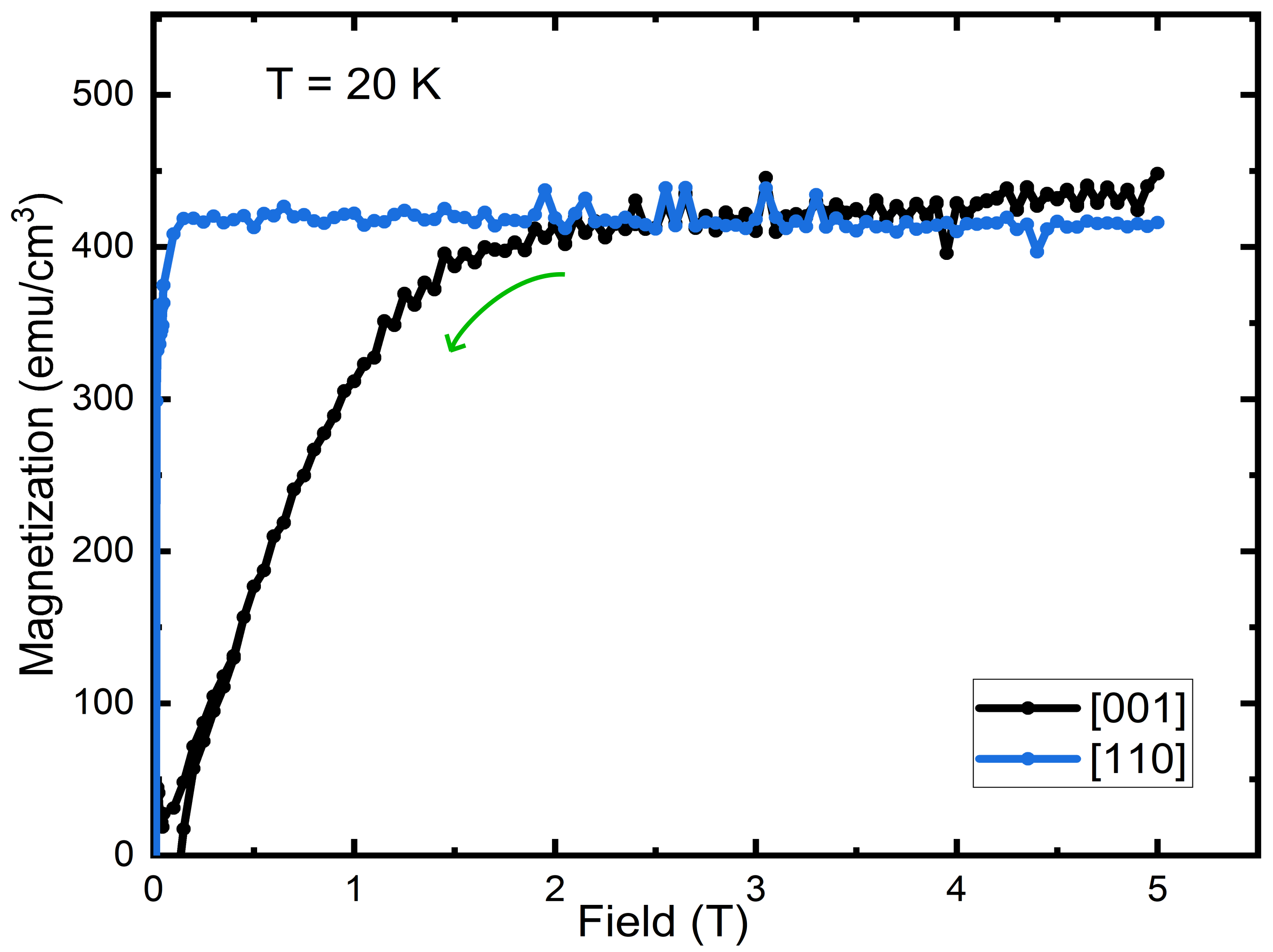}
\caption{Magnetization in an applied field up to 5~T, along [110] and [001]
  directions. Green arrows indicate the direction of the measurement.}
\label{squid_5T}
\end{figure}

The magnetization curves shown in Figure~\ref{squid_5T} are for a field applied along [110] and [001] film directions, after subtracting the substrate contribution (i.e. data collected at 300~K). It is clear, from the 'plateau' shape of the magnetization above 2~T, that the moments are saturated in this field.

\clearpage

%\section*{Bibliography}
\bibliographystyle{unsrt}
\bibliography{Supp-LSMO}